\documentclass[pra, a4paper, twocolumn, 10pt,showpacs]{revtex4}%
\usepackage{bbm, amsmath, amssymb, amsthm, bm,textcomp, nicefrac}
\usepackage[T1]{fontenc}
\usepackage[latin9]{inputenc}
\usepackage{graphicx}
\usepackage{geometry}

\newcommand{\ket}[1]{\left|{#1}\right\rangle}
\newcommand{\bra}[1]{\left\langle{#1}\right|}
\newcommand{\braket}[2]{\langle{#1}|{#2}\rangle}

\newcommand{\ketbrad}[1]{\left|{#1}\rangle\!\langle{#1}\right|}
\newcommand{\ketbra}[2]{\left|{#1}\rangle\!\langle{#2}\right|}

\newcommand{\C}{\ensuremath{\mathbbm C}}
\newcommand{\be}{\begin{equation}}
\newcommand{\ee}{\end{equation}}
\newcommand{\eea}{\end{eqnarray}}
\newcommand{\bea}{\begin{eqnarray}}
\newcommand{\Def}{ \mathrel{\mathop:}=}
\begin{document}

\title{Stability of encoded macroscopic quantum superpositions}

\author{F.\ Fr\"owis and W.\ D\"ur}

\affiliation{Institut f\"ur Theoretische Physik, Universit\"at
  Innsbruck, Technikerstr. 25a, A-6020 Innsbruck,  Austria}
\date{\today}

\begin{abstract}
  The multipartite Greenberger-Horne-Zeilinger (GHZ) state is a paradigmatic example of a highly entangled multipartite state with distinct quantum features. However, the GHZ state is very sensitive to generic decoherence processes, where its quantum features and in particular its entanglement diminish rapidly, thereby hindering possible practical applications. In this paper, we discuss GHZ-like quantum states with a block-local structure and show that they exhibit a drastically increased stability against noise for certain choices of block-encoding, thereby extending results of Ref.~[Phys. Rev. Lett. \textbf{106}, 110402 (2011)]. We analyze in detail the decay of the interference terms, the entanglement in terms of distillable entanglement, and negativity as well as the notion of macroscopicity as measured by the so-called index $q$, and provide general bounds on these quantities. We focus on an encoding where logical qubits are themselves encoded as GHZ states, which leads to so-called concatenated GHZ (C-GHZ) states. We compare the stability of C-GHZ states with other types of encodings, thereby showing the superior stability of the C-GHZ states. Analytic results are complemented by numerical studies, where tensor network techniques are used to investigate the quantum properties of multipartite entangled states under the influence of decoherence.
\end{abstract}

\pacs{03.67.-a,03.65.Ud,03.65.Yz}

\maketitle

\section{Introduction}
\label{sec:introduction}

The theory of quantum mechanics offers an improvement of current technologies; namely, secure communication due to its inherent probabilistic character and a polynomial or exponential speedup for the computation of certain tasks \cite{NiCh00} as well as an enhanced sensitivity in metrology \cite{metrology}. The notion of entanglement is central to many of these applications, and provides a crucial distinction between classical and quantum devices. In order to use entanglement for practical purposes, quantum mechanical systems have to be scalable in such a way that their quantum features are preserved. However, larger quantum systems interact with their surroundings more intensely, which rapidly destroys quantum correlations within the system. Therefore, one of the main goals is to protect quantum information processes against influences from the uncontrollable environment. One way out of these difficulties is quantum error correction \cite{reviewQEC,St98,Go97}. This theory aims to correct errors induced by environmental noise by using redundant encodings together with measurements and manipulations in such a way that quantum correlations are preserved. However, active error correction is itself a difficult task and extremely good control over the quantum system is required.

Besides the technological interest in quantum mechanics, more fundamental questions arise whenever we think about the axioms of quantum mechanics, which are to some extent in contrast to our daily experience. One may ask at which length or energy scales quantum mechanics is valid. If it is restricted to the microscopic realm, where does the transition to the classical world occur? Or is it more what many physicists believe; namely, that quantum mechanics is in principle also valid on ``human'' scales and above? Then decoherence may be the reason why we never experience interference effects of macroscopically superposed states. Although it is extremely challenging to give an ultimate answer to these questions, the successful demonstration of interference effects, not explainable by classical physics, confirms the validity of quantum mechanics on the scale of the experiment. To this end, the generation of quantum states that are well protected against noise is a central topic.

A very important quantum state for both issues is the so-called multipartite Greenberger-Horne-Zeilinger (GHZ) state
\begin{equation}
\label{eq:38}
\ket{\mathrm{GHZ}_N} = \frac{1}{\sqrt{2}}\left( \ket{0}^{\otimes N} + \ket{1}^{\otimes N}  \right)
\end{equation}
(for the notation, see Sec.~\ref{sec:notations-setting}). On the one hand, it is a useful resource for different applications like certain security tasks in distributed communication scenarios \cite{multiQKD} or quantum metrology \cite{Huelga}. In addition, the GHZ state reflects to some extent the idea of Schr\"odinger's cat \cite{Schroedinger35} (i.e., a ``macroscopic superposition'' of two ``classical'' states). For a small number of particles, the GHZ state has already been demonstrated experimentally \cite{Experiments,IonBlatts}. On the other hand, several studies such as Refs.~\cite{aolita,simon,ClusterStable,ConcatEC,Huelga} show that the GHZ state is also very sensitive to the effect of noise and decoherence. For generic decoherence processes, practically all quantum properties vanish exponentially fast with the number $N$ of particles. It is therefore of fundamental interest to search for alternative quantum states that show similar properties and --at the same time-- are more stable against loss of coherence and entanglement due to environmental influence. This would open the way towards the generation of large-scale entanglement and the observation of the corresponding quantum effects in mesoscopic or even macroscopic systems.

\subsection{Basic idea, setting and notation}
\label{sec:notations-setting}

In this section we formulate the main idea to increase the stability of GHZ-like states, thereby reviewing and extending results of Ref. \cite{cGHZ}. The setting is the Hilbert space $\mathcal{H}$ of $m\times N$ two-level systems, called qubits, each defined on $\C^2$ ($m$ and $N$ are integers). The complete space is therefore $\mathcal{H} = \C^{2\otimes (mN)}$. In the theory of quantum error correction it is common to group the qubits in order to distribute the basic information unit over several physical units. To this end, two orthogonal, \textit{logical} states are defined, hereafter denoted by $\ket{\tilde{0}_L}, \ket{\tilde{1}_{L}} \in \C^{2\otimes m}$. A logical GHZ state is then defined as
\begin{equation}\label{eq:2a}
\ket{\mathrm{GHZ}_L} = \frac{1}{\sqrt{2}}\left(\ket{\tilde{0}_L}^{\otimes N}+ \ket{\tilde{1}_L}^{\otimes N}\right).
 \end{equation}
With error correction (i.e., active measurements and operations on this block), it is possible to protect the quantum information to some extent against uncontrollable influence from the environment \cite{NiCh00,reviewQEC,St98,Go97}.

Error correction is indeed a successful strategy to accomplish this task. However, the required coding and decoding operations are often rather complex, and a successful protection requires repeated active intervention (i.e., error syndrome readout and correction operations). Error correction codes are designed such that they can fully protect arbitrary quantum information against the influence of local noise and imperfections, as long as the noise is sufficiently small or errors appear only on some of the particles where information is encoded. In many cases the relevant criteria to identify good codes is the number of acceptable errors in relation to the size of the codewords. For instance, the optimal code capable of correcting an arbitrary single qubit error is of size five [e.g., a 5-qubit Calderbank-Shor-Steane (CSS) code \cite{5QubitCss}]. However, in order that error correction be successful, only a small amount of noise is acceptable, requiring repeated active intervention at short time intervals. We remark that there also exist codes specially designed to cope with a large amount of noise \cite{DiVincenzo98}, where general local errors of about 19\% can be successfully corrected. Clearly, also without active intervention, error-correction codes offer some protection of encoded quantum information and certain quantum features such as entanglement. The same is true in situations when the noise is larger than what the code can successfully correct. In general, however, quantum error correction codes are not designed or optimized for such scenarios.

In this work we also form $N$ groups, each consisting of $m$ qubits. However, the main idea here is to find an encoding so that the quantum information is protected without any active intervention. A very special encoding was introduced in \cite{cGHZ}. Given an orthonormal basis $\ket{0}$ and $\ket{1} \in \mathbbm{C}^2$ (here the eigenstates of the Pauli matrix $\sigma_z$) we define
\begin{equation}
  \label{eq:1}
  \begin{split}
    \ket{0_L} &= \frac{1}{\sqrt{2}}\left(\ket{0}^{\otimes m}+ \ket{1}^{\otimes
        m}\right),\\
    \ket{1_L} &= \frac{1}{\sqrt{2}}\left(\ket{0}^{\otimes m}- \ket{1}^{\otimes
        m}\right).
  \end{split}
\end{equation}
Note that we treat the logical blocks as being local; that is, we consider higher dimensional local systems. Whenever in the course of this paper we perform ``local'' operations, we refer to actions on the whole block. 

We also remark that the encoding is similar to the Shor code \cite{ShorCode}. That is, for $m=3$ the concatenated GHZ (C-GHZ) state is equivalent to a logical GHZ of size $N/3$ using the Shor code with blocks of nine qubits. This is, however, not the setting we consider here.  In fact, the code we use is a simple repetition code that --when used for error corrections-- cannot correct for arbitrary errors. However, as we will show in the following, several quantum features of the encoded states are nevertheless well preserved even without active error correction.

In \cite{cGHZ} we used the so-called GHZ encoding (\ref{eq:1}) to demonstrate the enhanced stability of the logical GHZ state (\ref{eq:2a}). The resulting state \begin{equation}
  \label{eq:2}
\ket{\phi_c} = \frac{1}{\sqrt{2}}\left(\ket{0_L}^{\otimes N}+ \ket{1_L}^{\otimes N}\right)
  \end{equation}
is the C-GHZ state. In the following, we will review and extend the results of \cite{cGHZ}.

We investigate several quantum properties of the C-GHZ state under the influence of noise. Here, we focus on decoherence models which are describable as completely-positive trace-preserving maps (cp maps). Furthermore, for the remainder of this article, only uncorrelated decoherence is regarded (i.e., every qubit interacts with its individual environment). If we denote the cp map acting on qubits $i$ by $\mathcal{E}_t^{(i)}$, the overall decoherence process is written as $\mathcal{E}_t = \bigotimes_{i=1}^{mN} \mathcal{E}_t^{(i)}$. The parameter $t$ refers to the elapsing time and we assume that, for $t=0$, the mapping reduces to the identity $\mathcal{E}_0(\rho) = \rho$. We use a certain representation in the Pauli basis $\{\sigma_i\} (i=0,\dots,3)$, where $\sigma_0$ refers to the identity operator. Every single-qubit cp map can therefore be written as
  \begin{equation}
    \label{eq:3}
    \mathcal{E}^{(i)}_t(\rho) = \sum_{k,l=0}^3 \lambda_{kl}(t)\sigma_k^{(i)}\rho\sigma_l^{(j)}.
  \end{equation}
  The coefficients $\lambda_{kl}(t)$ have to be such that the action of $\mathcal{E}^{(i)}_t$ even on parts of the system maps any density matrix to a density matrix.

  For instance, the cp map with $\lambda_{kl} =0$ for all $k\neq l$ is called Pauli noise. Here we mainly concentrate on the depolarization channel (white noise) $\mathcal{D}_{t}$ which takes the form
  \begin{equation}
    \label{eq:5}
    \mathcal{D}^{(i)}_t(\rho) = p(t)\rho + \frac{1-p(t)}{4}\sum_{k=0}^3\sigma_k^{(i)}\rho\sigma_k^{(i)},
  \end{equation}
 with the noise parameter $p(t) = e^{-\gamma t}$, where $\gamma$ is the decoherence rate. An alternative description of this noise
  \begin{equation}
    \label{eq:5a}
    \mathcal{D}^{(i)}_t(\rho) = p(t)\rho + \frac{1-p(t)}{2} \mathrm{Tr}_i(\rho) \otimes \mathbbm{1}^{(i)}
\end{equation}
explains the name white noise, since in the long-time limit $\mathcal{D}_t$ transforms every quantum state to the complete mixture. This mapping models the system in contact with a thermal bath at infinite temperature. We see that this decoherence model is basis independent and there do not exist any decoherence free subspaces. For this reason we focus in this article on uncorrelated white noise, hence the noisy quantum states of Eq.~(\ref{eq:2}) read
\begin{equation}
\label{eq:39}
\begin{split}
  \rho_c&= \mathcal{D}_t^{(1)}
  \mathcal{D}_t^{(2)}\dots \mathcal{D}_t^{(mN)}\ketbrad{\phi_c}\\ &\equiv\mathcal{D}_t\ketbrad{\phi_c}
\end{split}
\end{equation}
Notice that we will denote white noise processes acting on individual qubits by $\mathcal{D}_t$ or $\mathcal{D}$ throughout this article, where the number of particles the process acts on is implicitly given by the size of the state.

Although one can argue \cite{standardForms} that, to some extent, the depolarization channel represents a ``worst case scenario'', in Sec. \ref{sec:diagonal-elements} we will explain why our findings are qualitatively valid for any time-dependent noise model of the form (\ref{eq:3}).

We remark that noise processes with a preferred basis are, in general, less harmful, as a local basis change can significantly reduce the influence of such noise processes on specific states, like the GHZ state, as pointed out in Ref.~\cite{Chaves11}. Such an observation, however, only holds if some kind of noise is absent or strongly suppressed.  If one considers, for example, Pauli noise, the smallest coefficient $\lambda_{k}$ determines the level of white noise, as one may view Pauli noise as a white noise process plus some additional noise.

Here, we are interested in a stability against arbitrary noise processes. However, in the presence of noise processes with preferred basis or correlated decoherence, additional stability can be obtained by the above-mentioned techniques \cite{Chaves11} or decoherence-free subspaces, respectively.

\subsection{Summary of results and outline}
\label{sec:summary-results}

Here, we provide a short summary of the results of the paper. All expressions used here are defined properly in the subsequent sections.

The paper critically reviews properties of encoded GHZ-type states, in particular the C-GHZ states, when subjected to decoherence described by uncorrelated depolarization channels [single-qubit white noise described by parameter $p$; see Eq.~(\ref{eq:39})]. We establish not only generally valid bounds for the decay of coherence, the decay of entanglement (measured by negativity), and the lifetime of distillable entanglement for arbitrary encodings, but also provide explicit results for C-GHZ states (see also Ref.~\cite{cGHZ}).

The first results are presented for the decay of the so-called interference terms (off-diagonal elements), which are those terms of the density operators that manifest the difference to an incoherent mixture of classical states. Section \ref{sec:diagonal-elements} motivates these considerations, and shows explicit formulas in the case of the trace norm. First we show a general upper bound on the stability of coherence (off-diagonal elements) for GHZ-type states with arbitrary block-wise encoding, which demonstrates that an exponential decay with system size $N$ is unavoidable but is (exponentially) slowed down by considering blocks of size $m$. For the C-GHZ state we find that the off-diagonal elements of Eq.~(\ref{eq:39}) decay in fact drastically slower compared to the standard GHZ state and therefore the coherences are stabilized for a certain time interval. We also find that the stability for C-GHZ states is very close to the general upper bound, thereby indicating optimality. For sufficiently large $p$ (i.e.~for small interaction times $t$) we establish a lower bound on the decay of off-diagonal elements for the C-GHZ state and show that this lower bound tends to unity in the limit of large system sizes $N$ under the condition that the block size $m$ of a logical cell grows logarithmically with $N$. This shows that one can stabilize the decay of the off-diagonal elements for arbitrary system sizes. The remaining paragraphs of this section deal with general arguments why this stability is generically valid, irrespective of the actual decoherence channel. Furthermore,k we mention shortly the behavior for another choice of norm, the Hilbert-Schmidt norm. That is mainly interesting for comparisons with other types of encodings in Sec.~\ref{sec:alternatives}, where the calculation of the trace norm is not feasible.

An important issue in quantum information theory is addressed in Sec.~\ref{sec:entangl-prop}: entanglement. The ability to distill multipartite entanglement out of a large number of noisy quantum states is demonstrated for the C-GHZ state. It is shown that increasing the group size $m$ leads to exponential growth of the maximal possible system size $N$ such that $N$-party entanglement can be generated. Again, a logarithmically growing block size $m = \log N$ suffices to ensure distillability for arbitrary finite $N$. Similar results are found for a computable measure of bipartite entanglement, the so-called negativity. For the standard GHZ state, the decay of negativity with $N$ can be approximated by an exponential function. Also the C-GHZ states loose their negativity exponentially fast with $N$, but this decay rate can be decreased itself exponentially fast by increasing $m$. Hence, we also claim that the negativity can be stabilized by a suitable choice of the block size $m$. In addition, we establish a generally valid upper bound for the negativity depending exclusively on the trace norm of the off-diagonal elements. In contrast to these results, we find that the lifetime of genuine multipartite entanglement cannot be increased for the C-GHZ state.

As already mentioned, the GHZ state serves as an archetypal Schr\"odinger-cat state. Sometimes, it has therefore been called a macroscopic quantum superposition. The question is whether it still can be seen as such in the presence of a decoherence process.  In Sec.~\ref{sec:c-ghz-state} we adapt a measure for macroscopicity from the literature --the so-called index $q$-- and study its behavior under decoherence. Comparable to the negativity, we find that the index $q$ decays much slower for C-GHZ states with higher $m$-values. Also, here we are able to report on an upper bound for an assigned effective size as a function of the trace norm of the interference terms.

The question whether the encoding of Eq.~(\ref{eq:1}) discussed so far is the optimal one is investigated in Sec.~\ref{sec:alternatives}. Instead of using GHZ states as codewords [cf.~Eq.~(\ref{eq:1})], we use one-dimensional (1D) cluster states \cite{ClusterState}. A 1D cluster state of five qubits with periodic boundary conditions can be used as a codeword for quantum error correction, and we compare the stability of resulting logical GHZ states using such an encoding with and without active error correction to the GHZ encoding used in the C-GHZ state. Furthermore, we also investigate a state vector called a cluster-GHZ state, a superposition of two orthogonal, $N$ qubit 1D cluster states, and compare such as state with the C-GHZ state. This is done using techniques from the field of tensor network states, which is interesting on its own, since it is used to represent cp maps and to calculate properties of quantum states in an numerically exact manner \cite{MPSO}. Later, the idea of using GHZ or cluster encoding is iterated, where we consider concatenated encodings and investigate the stability of the resulting states. Finally, we perform random searches for good codewords. The conclusion of this section is that GHZ encoding (\ref{eq:1}) is superior to all other encodings discussed here.

As a last point, we show in Sec.~\ref{sec:exper-real-ion} that a feasible and efficient generation of the C-GHZ states is possible using existing ion trap setups. In addition, we point out that systems of only a few ions already suffice to demonstrate the stabilization effect.


  \section{Off-diagonal elements}
  \label{sec:diagonal-elements}

  In this section, we consider the decay of the off-diagonal elements (or interference terms) under the depolarization channel $\mathcal{D}_t$. This is  quantified by calculating their trace norm. Motivating and discussing these investigations covers the main part of the present section. At the end, we also have a look at arbitrary cp maps and the behavior of the Hilbert-Schmidt norm of the interference terms.

  \subsection{Why are interference terms interesting?}
  \label{sec:why-are-interference}

  If we write the C-GHZ state (\ref{eq:2}) as a density matrix
  \begin{equation}
    \label{eq:6}
    \begin{split}
      \ketbrad{\phi_C} =& \frac{1}{2}\left(\ketbrad{0_L}^{\otimes N}+
        \ketbrad{1_L}^{\otimes N}\right.\\ & \left.+\ketbra{0_L}{1_L}^{\otimes
          N}+\ketbra{1_L}{0_L}^{\otimes N}\right),
    \end{split}
  \end{equation}
  we recognize that the last two terms are those that distinguish this state from the incoherent mixture $\frac{1}{2}\left(\ketbrad{0_L}+ \ketbrad{1_L}\right)$. All nonclassical effects between the blocks we might observe are due to these interference terms, often also called coherences. So it seems reasonable to regard the norm of the off-diagonal elements as an indicator of the stability of coherence under environmental interaction. This can be compared to the norm of the diagonal elements \cite{NoteConnectionMacroMeasure}.

  A convenient choice for the norm is the trace norm \cite{NoteTraceNorm}, not only because the trace norm of the diagonal elements is unity under any circumstances and therefore easily comparable, but also because the trace norm of the noisy interference terms serves as an upper bound on two quantum properties that we are going to discuss in the course of this article. For the specific shape of the logical GHZ state (\ref{eq:2a}), the trace norm of the off-diagonal elements is always larger than the negativity, an entanglement measure treated in Sec. \ref{sec:negat-as-meas}. Additionally, in Sec. \ref{sec:c-ghz-state} we investigate whether the C-GHZ state can be called macroscopic even under the influence of noise. Also in this context we encounter an upper bound established by the trace norm of the off-diagonal elements.

\subsection{Upper bound for off-diagonal norm}
\label{sec:upper-bound-diagonal}

We start by establishing an upper bound on the decay of coherences for encoded GHZ states (\ref{eq:2a}) with arbitrary encoding. We consider the trace norm of the off-diagonal elements under the influence of the depolarization channel
\begin{equation}
  \label{eq:7a}
J \equiv \lVert \mathcal{D}(\ketbra{\tilde{0}_L}{\tilde{1}_L}^{\otimes N})\rVert_{1} = \lVert \mathcal{D}(\ketbra{\tilde{0}_L}{\tilde{1}_L})\rVert_{1}^N \equiv (J_0)^N, \end{equation}
where $\mathcal{D}$ denotes individual white noise processes acting on all $Nm$ qubits, or the $m$ qubits of a logical block, respectively. For notational convenience we skip the index $t$ here. From this expression, it is already clear that an exponential decay with $N$ is unavoidable.

To calculate $J$, it is sufficient to evaluate $J_0$ (i.e., consider only one block of $m$ qubits). We consider two arbitrary orthogonal states $\ket{\tilde{0}_L}, \ket{\tilde{1}_L}$; in particular, the trace norm of the off-diagonal element $\mathcal{D}(\ketbra{\tilde{0}_L}{\tilde{1}_L})$. The action of white noise [see Eq.~(\ref{eq:5a})] to all $m$ qubits leads to
\begin{equation}
\begin{split}
  J_0=\mathcal{D}&(\ketbra{\tilde{0}_L}{\tilde{1}_L}) = p^m \ketbra{\tilde{0}_L}{\tilde{1}_L} \\ & + p^{m-1}\frac{1-p}{2} \sum_{i=1}^m \mathrm{Tr}_i(\ketbra{\tilde{0}_L}{\tilde{1}_L}) \otimes \mathbbm{1}^{(i)}  \\ & + p^{m-2}\left(\frac{1-p}{2}\right)^{2}
\sum_{i,j=1}^m \mathrm{Tr}_{i,j}(\ketbra{\tilde{0}_L}{\tilde{1}_L}) \otimes \mathbbm{1}^{(i,j)}  \\ &+ \dots +\left(\frac{1-p}{2}\right)^{m} \mathrm{Tr}(\ketbra{\tilde{0}_L}{\tilde{1}_L}) \mathbbm{1}^{\otimes m}.
\end{split}\label{eq:8}
\end{equation}
The last term in  Eq.~(\ref{eq:8}) vanishes, because $\braket{\tilde{0}_L}{\tilde{1}_L} = 0$. To estimate the trace norm of  Eq.~(\ref{eq:8}), we use the triangle inequality and recognize that $\lVert \mathrm{Tr}_i(Q)\rVert_1\leq \lVert Q\rVert_1$ for any linear operator $Q\in \C^{2\otimes m}$. Since $\lVert \ketbra{\tilde{0}_L}{\tilde{1}_L}\rVert_1 = 1$, the result reads
\begin{equation}
  \label{eq:9}
  \begin{split}
    \lVert \mathcal{D}(\ketbra{\tilde{0}_L}{\tilde{1}_L})\rVert_1 &\leq \sum_{i=0}^{m-1}\binom{m}{i}p^{m-i}(1-p)^i \\ &= 1-(1-p)^{m}.
  \end{split}
\end{equation}

We have therefore established an upper bound on the trace norm of any off-diagonal element of a single block $J_0=\lVert \mathcal{D}(\ketbra{\tilde{0}_L}{\tilde{1}_L})\rVert_1$. On the one hand, this again shows us that there does not exist any decoherence free subspace, because for any $t>0$ and finite $m$ the trace norm of the off-diagonal elements is strictly smaller than one. On the other hand, a large $m$ takes this upper bound arbitrarily close to unity and it does so exponentially fast. In fact, for sufficiently large $m = O(\log N)$ one also finds that $J=J_0^N$; that is, the coherences of the encoded GHZ state, remain arbitrarily close to unity for some finite time interval.

We have performed a numerical maximization of $\lVert \mathcal{D}(\ketbra{\tilde{0}_L}{\tilde{1}_L})\rVert_1 $ for fixed $p$ and $m=2$ and find a maximal value $p$ which is less than $p(2-p)$, as set by  Eq.~(\ref{eq:9}). This indicates that the upper bound can in general not be saturated. However, we will see that our choice [cf.~Eq.~(\ref{eq:1})] for $\ket{\tilde{0}_L}$ and $\ket{\tilde{1}_L}$ exhibits a trace norm which comes very close to the derived upper bound.

We also remark that the derived upper bound on $J_0$ not only leads to an upper bound on $J$ for encoded GHZ states of the form  Eq.~(\ref{eq:2a}), but also for coherences of arbitrary superposition states of $N$ qubits, $|\psi\rangle = \tfrac{1}{\sqrt 2}(|\psi_0\rangle + |\psi_1\rangle)$, $\lVert \mathcal{D}(\ketbra{\psi_0}{\psi_1})\rVert_1 \leq  1-(1-p)^{N}$.

\subsection{Results for the C-GHZ under white noise}
\label{sec:results-c-ghz}

In this paragraph we concentrate on the calculation of $J_0$ for a noisy C-GHZ state  Eq.~(\ref{eq:2}), where $|0_L\rangle, |1_L\rangle$ are given by  Eq.~(\ref{eq:1}) (see also Ref.~\cite{cGHZ}). To avoid confusion, we denote this quantity by $I_0$
\begin{equation}
  \label{eq:7}
I \equiv (I_0)^N = \lVert \mathcal{D}(\ketbra{0_L}{1_L})\rVert_{1}^N. \end{equation}
We first write the off-diagonal element in terms of the local standard basis $\ketbra{0_L}{1_L} = 1/2\left(\ketbrad{0}^{\otimes m}-\ketbra{0}{1}^{\otimes m}+\ketbra{1}{0}^{\otimes m}- \ketbrad{1}^{\otimes m}\right)$. The depolarization channel alters the first and the last term by distributing their amplitudes over the whole diagonal in this representation. These elements themselves therefore keep their unit trace norm. In contrast, the second and the third elements are purely damped by a strong factor $p^m$. In total, the interference term under white noise reads
\begin{equation}
  \label{eq:10}
  \begin{split}
   \mathcal{D}(\ketbra{0_L}{1_L})&=    \sum_{k_1,\dots,k_m=0}^1 c_{s}^{-}    \ketbrad{k_1\dots k_m}\\ &-p^m/2 \ketbra{0}{1}^{\otimes m}+p^m/2 \ketbra{1}{0}^{\otimes m},
  \end{split}
\end{equation}
where $s =\sum_i k_i$ is the number of $\ket{1}$-states in the basis vector $\ket{k_1\dots k_m}$. We use
\begin{equation}
\label{eq:45}
c_{s}^{\pm} = \frac{1}{2^{m+1}} \left[(1+p)^{m-s}(1-p)^s \pm (1+p)^{s}(1-p)^{m-s}\right]
\end{equation}
which depends only on $s$. The calculation of $I_0$ is simple, because  Eq.~(\ref{eq:10}) is almost diagonal. We arrive at $I_0 =  \sum_{s = 0}^m \binom{m}{s} \lvert c_s^{-}\rvert$. The coefficients $c_s^{-}$ are positive semidefinite as long as $s\leq \lfloor m/2\rfloor$. If we use $\sum_{s = 0}^{\lfloor m/2 \rfloor}\binom{m}{s}(1+p)^{s}(1-p)^{m-s} = \sum_{s = \lceil m/2\rceil}^m \binom{m}{s}(1+p)^{m-s}(1-p)^s$ and $\frac{1}{2^{m}}\sum_{s = 0}^{m}\binom{m}{s}(1+p)^{s}(1-p)^{m-s} = 1$, we achieve the final result for even $m$
\begin{equation}
  \label{eq:11}
  \begin{split}
    I_0 &= 1-\frac{1}{2^{m}}
    \binom{m}{m/2}(1-p^2)^{m/2}\\ &-\frac{1}{2^{m-1}}\sum_{s = m/2+1}^m
    \binom{m}{s}(1+p)^{m-s}(1-p)^s,
  \end{split}
\end{equation}
and for odd $m$
\begin{equation}
  \label{eq:11a}
I_0 =  1-\frac{1}{2^{m-1}}\sum_{s = \lceil m/2\rceil}^m \binom{m}{s}(1+p)^{m-s}(1-p)^s.
\end{equation}

\begin{figure}[htbp]
\centerline{\includegraphics[width = 1\columnwidth]{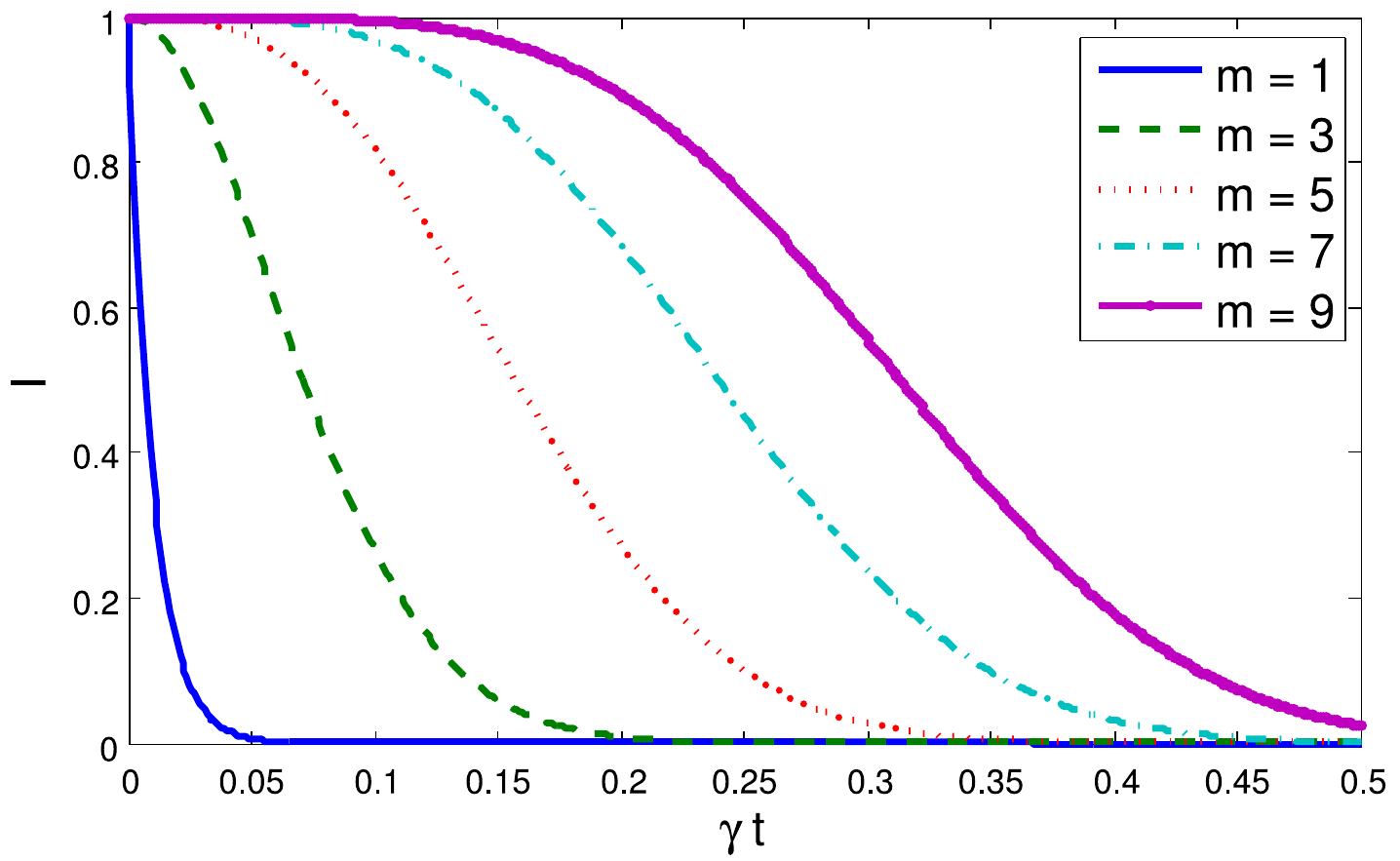}}
\caption[]{\label{fig:tracenorm} (Color online) Trace norm of off-diagonal element $I$ for $N=100$ in units of the dimensionless time $\gamma t$. The exponentially fast stabilization effect can clearly be seen.}
\end{figure}
A numerical instance for fixed $N$ and several $m$ can be found in Fig.~\ref{fig:tracenorm}. What follows is a detailed discussion of $I_0$. In order to make it easier, we concentrate on odd $m$. The conclusions are exactly the same for both cases, only the detailed formula differ slightly. Notice that, for even $m$, Eq.~(\ref{eq:11}) provides the same expression as  Eq.~(\ref{eq:11a}) for $m-1$.

First we observe that, since the sum in Eq.~(\ref{eq:11a}) starts from $s = \lceil m/2\rceil$, the first $\lfloor m/2 \rfloor$ derivatives of $I_0$ vanish at $t = 0$. This can also be seen if one exchanges $p$ in  Eq.~(\ref{eq:11a}) by the Taylor series around $t=0$. Then we have $I_0 = 1- \binom{m}{\lceil m/2 \rceil}/ 2^{m-1}\, (\gamma t)^{\lceil m/2 \rceil} + O((\gamma t)^{\lceil m/2 \rceil+1})$. This shows that for small times the off-diagonal norm is exponentially stabilized with $m$.

Next we consider the full off-diagonal $I$ of  Eq.~(\ref{eq:7}). We again simplify the discussion by working with a lower bound on $I_0$. To this end, we notice that the first addend in the sum of  Eq.~(\ref{eq:11a}) is always larger than the others, hence we have $I_0 \geq 1-2^{-m}\left(m-1\right) \binom{m}{\lfloor m/2 \rfloor} \left(1+p \right)^{(m-1)/2}\left(1-p \right)^{(m+1)/2}$. Applying Stirlings relation $\sqrt{2\pi n}\left(\nicefrac{n}{e}\right)^n < n! < \left(1+\nicefrac{1}{11n} \right)\sqrt{2\pi n}\left(\nicefrac{n}{e}\right)^n$ and doing some further simplifications we estimate
\begin{equation}
  \label{eq:13}
I \geq \left[1-\sqrt{\frac{2m}{\pi}}\left(1+\frac{1}{11m} \right)\left(1-p^2 \right)^{\lceil m/2 \rceil}\right]^{N}.
\end{equation}
It is clear that, for fixed $m$, Eq.~(\ref{eq:13}) goes down exponentially fast with $N$. But we can stabilize $I$ if $m$ is allowed to grow logarithmically with $N$ (i.e., we set $N = b^m$, $b>1$). For small times we approximate $I \gtrapprox 1- b^m \sqrt{\frac{2m}{\pi}}\left(1+\frac{1}{11m} \right)\left(1-p^2 \right)^{\lceil m/2 \rceil}$ and see that this tends to one in the limit of large $m$ as long as $b\sqrt{1-p^2}<1$. For instance, the choice $b=2$ (i.e.~$m = \log_2 N$) limits the validity of this approximation to $p>\sqrt{3}/2\approx 0.85$. In summary, a lower bound derived for the trace norm of the interference term tends to one in the limit of large system sizes, provided that $m$ grows logarithmically with $N$.

\begin{figure}[htbp]
\centerline{\includegraphics[width = 1\columnwidth]{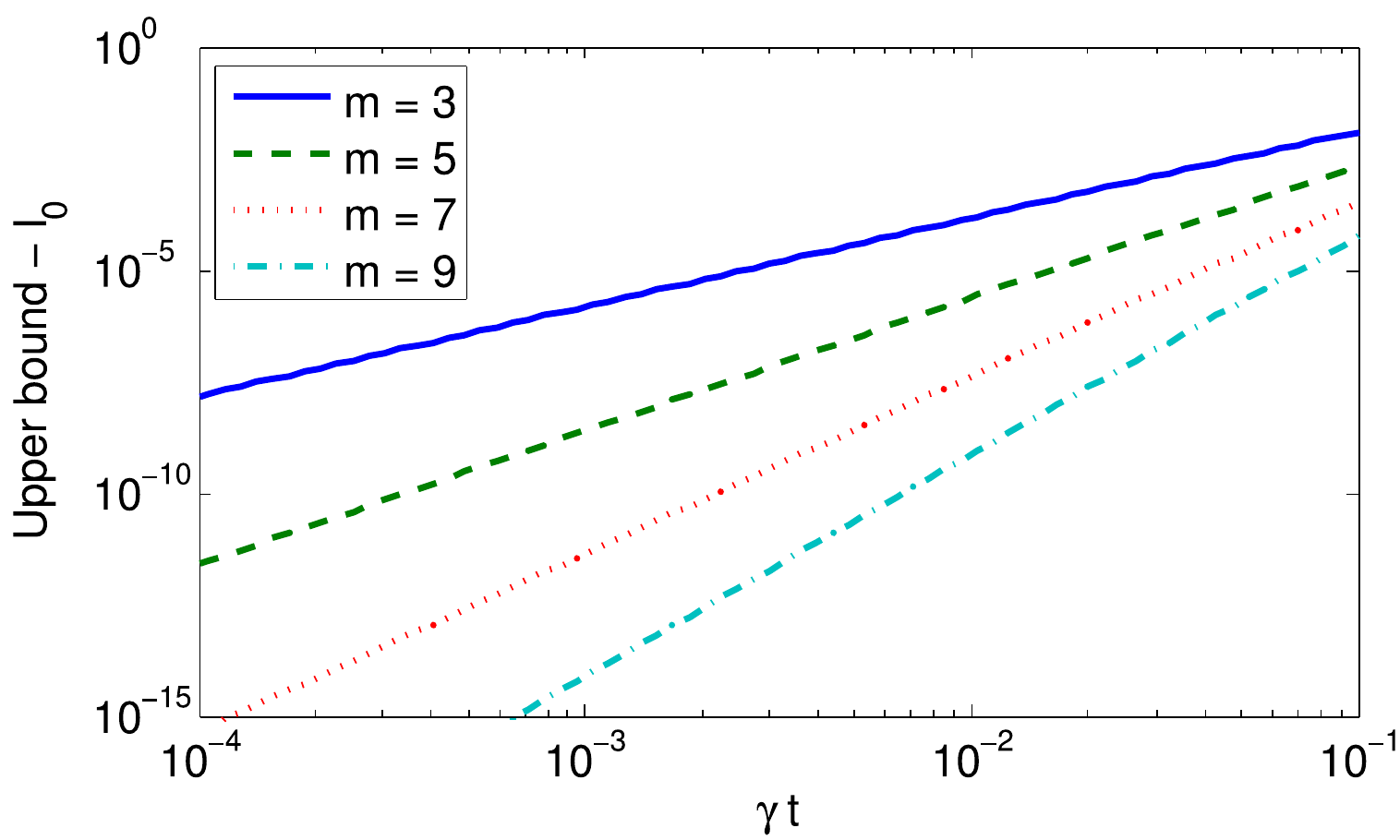}}
\caption[]{\label{fig:Diff-TraceNorm-UpperBound} (Color online) Absolute difference between  upper bound of Eq.~(\ref{eq:9}) and trace norm of off-diagonal element (\ref{eq:11a}) are shown for several choices of $m$. One observes that $I_0$ approaches the upper bound for larger $m$ and small times $t$.}
\end{figure}

Finally, we would like to stress that, for small times, $I_0$ comes very close to the ultimate bound for the trace norm of off-diagonal elements as derived in Sec. \ref{sec:upper-bound-diagonal}, which is $I_0\leq 1- \left(1-p\right)^m$. This can be seen most easily by inspecting Fig.~\ref{fig:Diff-TraceNorm-UpperBound} where we see numerical examples for the difference between $I_0$ and the established upper bound  Eq.~(\ref{eq:9}) for different block sizes $m$. Notice that $I_0$ approaches the upper bound for small times and increasing $m$.

This concludes the discussion of the trace norm of the interference elements under the effect of white noise. We have shown that increasing the logical block size $m$ stabilizes $I_0$ and therefore $I$ exponentially fast.

\subsection{Trace norm under general noise models}
\label{sec:trace-norm-under}

The depolarization channel is a very effective decoherence process, since it affects all bases. Nevertheless it is an interesting question to ask how the off-diagonal elements of the C-GHZ state behave under other noise models. One can easily calculate the trace norm for other important instances as the damping channel, dephasing noise or the phase damping. The result is essentially the same as for the depolarization channel; namely, that the decay of the off-diagonal could always be stabilized exponentially fast. Instead of listing all the results of the above examples we would rather like to give a general argument why, for small times, the interference terms of the C-GHZ state always exhibit this stability for every decoherence process that can be described as an uncorrelated cp map $\mathcal{E}_t$. We remark that for noise with a preferred basis, additional stabilization is possible by adapting the local basis to the specific noise process (see, e.g., Ref.~\cite{Chaves11}).

Before we start, we fix the parametrization of the cp map. We consider any physically meaningful model that starts the decoherence process at $t=0$ and is continuous. In this case, a general cp map of the form of Eq.~(\ref{eq:3}) can be approximated for small times by setting $\lambda_{00}(t) = 1- \mu_{00}t + O(t^2)$ and $\lambda_{ij}(t) = \mu_{ij}t + O(t^2)$ for all other coefficients.

The trace of the expression
\begin{equation}
  \begin{split}
    \mathcal{E}_t\left(\ketbra{0_L}{1_L}\right) &=   \frac{1}{2}\left[\mathcal{E}_t\left(\ketbrad{0}\right)^{\otimes m} -\mathcal{E}_t\left(\ketbra{0}{1}\right)^{\otimes m} \right.\\ &\left.+ \mathcal{E}_t\left(\ketbra{1}{0}\right)^{\otimes m} -\mathcal{E}_t\left(\ketbrad{1}\right)^{\otimes m}  \right]
  \end{split}
\label{eq:40}
\end{equation}
equals zero at any time. This is due to the canceling of the first and the last term on the right-hand-side (rhs) in Eq.~(\ref{eq:40}). These are the terms we want to concentrate on. The second and third are of little importance for this discussion. We will neglect them from now on until the end of this paragraph. We plug in our first-order expansion from above. If we expand the standard representation of $\mathcal{E}_t\left(\ketbrad{0}^{\otimes m}- \ketbrad{1}^{\otimes m}\right) \equiv \bigotimes_{i=1}^n \mathcal{E}_t^{(i)}\left(\ketbrad{0}^{\otimes m}- \ketbrad{1}^{\otimes m}\right)$ to the order of $\lfloor m/2\rfloor$ errors or equivalently to $O(t^{\lfloor m/2\rfloor})$, we observe a block structure. The elements coming from $\ketbrad{0}^{\otimes m}$ do not mix with those coming from $\ketbrad{1}^{\otimes m}$. That means that in a series expansion of $\mathcal{E}_t\left(\ketbrad{0}^{\otimes m}- \ketbrad{1}^{\otimes m}\right)$, $\mathcal{E}_t\left(\ketbrad{0}^{\otimes m}\right)$ and $\mathcal{E}_t\left(- \ketbrad{1}^{\otimes m}\right)$ stay two separate blocks with unit trace norms [up to the corrections of order $O(t^{\lceil m/2 \rceil})$]. Therefore, the total trace norm is of the form $1-O(t^{\lceil m/2 \rceil})$, the same result as we had for the depolarization channel. As a last step, we notice that the norm of a matrix does not decrease by adding further blocks which appear by considering $\mathcal{E}_t\left(\ketbra{0}{1}^{\otimes m}\right)$ and $\mathcal{E}_t\left(\ketbra{1}{0}^{\otimes m}\right)$.

We now provide an alternative argument that shows that depolarizing (white) noise actually corresponds to a worst-case scenario among all possible local noise processes. Consider an arbitrary local noise process described by  Eq.~(\ref{eq:3}). The parameter $\lambda_{00}$ describes the error-free part of the process; namely, the Jamiolokowski fidelity of ${\cal E}$ with respect to the identity operation is given by $F= \lambda_{00}$ \cite{standardForms}. As shown in Ref.~\cite{standardForms}, one can map an arbitrary single-qubit noise process with given $\lambda_{00}$ to a depolarizing white noise process with the same $\lambda_{00}$. This is done by performing a random basis change before subjecting the system to the decoherence process, and undoing this random basis change at the end of the process. By this active change of basis, the system hence behaves as under the action of white noise, regardless of the actual form of the noise process. That is, stability under white noise implies stability under arbitrary (local) noise processes.

\subsection{Hilbert-Schmidt norm}
\label{sec:hilbert-schmidt-norm}

So far we focused on the trace norm to characterize the decay of the interference term under the influence of noise. The trace norm has an important meaning in the context of distance measures and the fidelity of quantum states, but sometimes it is hard to calculate. In contrast, the Hilbert-Schmidt norm (or two-norm) $\lVert A \rVert_2 = \sqrt{\mathrm{Tr}A A^{\dagger}}$ can be calculated easily. In this section we will shortly discuss the behavior of the Hilbert-Schmidt (HS) norm of the off-diagonal elements. We compare those with the decay of the diagonal, since also the diagonal elements do not exhibit a constant two-norm. In general, the HS norm of the diagonal decays faster for larger systems.

Again, it is sufficient to consider a single logical block of $m$ qubits, as the results for the logical GHZ state can be obtained by taking the $N^{\rm th}$ power of the results for a single block. The ratio $\tilde{I}_0=\lVert \mathcal{D}(\ketbra{0_L}{1_L})\rVert_2/\lVert\mathcal{D}(\ketbrad{0_L})\rVert_2$ can be calculated straightforwardly. We find
\begin{equation}
\label{eq:35}
\begin{split}
  \tilde{I}_0&=\sqrt{\frac{\sum_{s=0}^m \binom{m}{s}c_s^{- 2}
      +p^{2m}/2}{\sum_{s=0}^m \binom{m}{s}c_s^{+ 2} +p^{2m}/2}}\\
  &= \sqrt{\frac{\left( 1+p^2 \right)^m-\left( 1-p^2 \right)^m+\left( 2p^2 \right)^m}{\left( 1+p^2 \right)^m+\left( 1-p^2 \right)^m+\left( 2p^2 \right)^m}}
\end{split}
\end{equation}
While the numerator of this expression alone decay faster with increasing $m$, $\tilde{I}_0$ itself behaves similarly to the trace norm $I_0$. One can see that the $l\mathrm{th} (l<m)$ derivative of the numerator and the denominator alone are identically at $p=1$. We can convince ourselves easily that this leads to the observation that the first $m-1$ derivatives of $\tilde{I}_0$ vanish at $p=1$. This shows the stability of $\tilde{I}_0$ for small times, similarly to $I_0$. The proof is based on induction. We abbreviate the numerator by $X$ and the denominator by $Y$, hence $\tilde{I}_0=X/Y$, and see that we can always write any derivative $i$ of $\tilde{I}_0$ as $\tilde{I}_0^{(i)}=\Omega_i/Y^k$ with $\Omega_i$ a function of $X,Y$ and their derivatives up to degree $i$ and some integer $k$. If we assume that $\tilde{I}_0^{(i)}|_{p=1} =0$, it implies that $\Omega_i|_{p=1}=0$. If a further derivation $\tilde{I}_0^{(i+1)}$ vanishes as well for $p=1$, it must be true that $\Omega_i' |_{p=1}=0$. Indeed it is, given a further assumption; namely, that $\Omega_i|_{p=1}=0$ because $\Omega_i$ consists of products of powers of $X, Y$ and derivatives (up to degree $i$) thereof so that, at $p=1$, they sum up to one. This is due to $X^{(l)}|_{p=1}=Y^{(l)}|_{p=1}$ for $l<m$. Then one can show that $\Omega_i'$ exhibits the same structure and hence vanishes at $p=1$ as well if $i+1<m$. The starting point is $\Omega_1 = X' Y - X Y'$. Therefore, the ratio $\tilde{I}_0$ can be stabilized with increasing $m$.

The big advantage of the relative HS norm is the easy calculation compared to the trace norm for more complex systems. In Sec.~\ref{sec:alternatives} the relative HS norm is discussed for other encoding schemes than  Eq.~(\ref{eq:1}). There we discuss quantum states where an efficient description in terms of matrix product states exists. These representations are most suitable to calculate the two norm.

\section{Entanglement properties}
\label{sec:entangl-prop}

In this section we investigate the entanglement properties of noisy C-GHZ states, thereby reviewing and extending the results of \cite{cGHZ}.

\subsection{Distillability of noisy C-GHZ state}
\label{sec:dist-prop-c}

The principle idea of distilling entanglement is to create a highly entangled state out of many copies of noisy entangled states in a probabilistic fashion by means of local operations and classical communication. In the multipartite case a sufficient strategy is to distill a maximally entangled state between any two parties. From this point, every desired state can be generated through local operations via teleportation \cite{Du00}. Local in our case means that we allow for operations on the predefined blocks of size $m$ --which constitute one of the $N$ parties-- but no joint operations between different blocks. That is, we consider $N$ parties each holding a $2^m$ dimensional systems (or several copies thereof) which they can locally access.

Let us first consider standard $N$-qubit GHZ states [Eq.~(\ref{eq:38})]. The lifetime of distillable entanglement of such states has been analyzed in detail in Ref.~\cite{ClusterStable,ConcatEC}, where upper- and lower bounds were derived. In particular, for any $N$-qubit GHZ state a threshold value on $p$ is found, below which $N$-party entanglement vanishes. In turn, for any given noise level $p$, there is a maximal system size $N$ such that the state is $N$-party entangled. These results have been extended to a situation where several qubits are grouped into blocks of size $m$, and one considers whether $N$-party entanglement is present in the system. This corresponds to encoded GHZ state  Eq.~(\ref{eq:2a}), with an encoding $|\tilde{0}_L\rangle=|0\rangle^{\otimes m}$ and $|\tilde{1}_L\rangle=|1\rangle^{\otimes m}$. Even if one allows for block sizes $m \to \infty$, one finds that, for a given noise level $p$, the maximum number of blocks that can be $N$-party entangled is bounded. A numerical example for $p=0.9$ results in a maximal system size of $N=53$. We conclude that any $Nm$-particle GHZ states  which is subjected to local white noise noise of at least this strength does not contain any multipartite entanglement between more than 53 parties, even if the individual blocks are of arbitrary size.

We now turn to the distillability properties of C-GHZ states. One possible strategy is to aim for generating bipartite entanglement between two parties by acting on a single copy of the multipartite state. That is, all but two parties perform local measurements on their systems. From the resulting two-party density operators one can then perform a standard bipartite distillation protocol in order to obtain a pure, maximally entangled state. Notice that for this step, several copies of the noisy bipartite states are required. Since the whole procedure is done by means of local operations and classical communication, the entanglement for these new states must have been supported by the original noisy states. Therefore such states contain distillable multipartite entanglement. This strategy is simple to analyze, as only the distillability of the resulting bipartite state needs to be checked. If each of the two parties projects their system onto a two-dimensional subspace [e.g., the one spanned by $|0_L\rangle,|1_L\rangle$; see  Eq.~(\ref{eq:1})], it is sufficient that the resulting density operator has negative partial transpose \cite{Du00}. A fidelity of $F > 1/2$ with a maximally entangled two-qubit state of this subspace, e.g. $|\mathrm{GHZ}_2\rangle =1/\sqrt{2}(|0_L0_L\rangle+ |1_L1_L\rangle)$, guarantees a negative partial transposition and ensures distillability. This supplies us with a lower bound on the lifetime of distillable multipartite entanglement.

In the following, we provide a simple explicit distillation protocol of this kind that shows that the use of the GHZ encoding [cf.~Eq.~(\ref{eq:1})] improves significantly the usability of logical GHZ states without any active error correction. This protocol consists of the following steps: 1. All but two logical blocks are subject to a local \cite{NoteLocalMeas} measurement. More specifically, we project onto the $\ket{0}^{\otimes m}$ --receiving the outcome $\lambda_{+}$-- or on the $\ket{1}^{\otimes m}$ state with outcome $\lambda_{-}$. 2. The remaining two blocks are also reduced to an effective dimension two by projecting onto the subspace spanned by $\{|0_L\rangle,|1_L\rangle\}$ or equivalently  $\{\ket{0}^{\otimes m}, \ket{1}^{\otimes m}\}$. 3. The distillability of the resulting (logical) two-qubit states is analyzed by calculating the overlap with a maximally entangled (logical) Bell state $\ket{\Phi^{\pm}_L} = 1/\sqrt{2}\left(\ket{0_L0_L}\pm\ket{1_L1_L}\right)$. The choice of the plus or the minus phase depends on the number of $\lambda_{-}$-outcomes, where an even [odd] number of $\lambda_{-}$ outcomes corresponds to a plus [minus] phase respectively.

\begin{figure}[htbp]
\centerline{\includegraphics[width = 1\columnwidth]{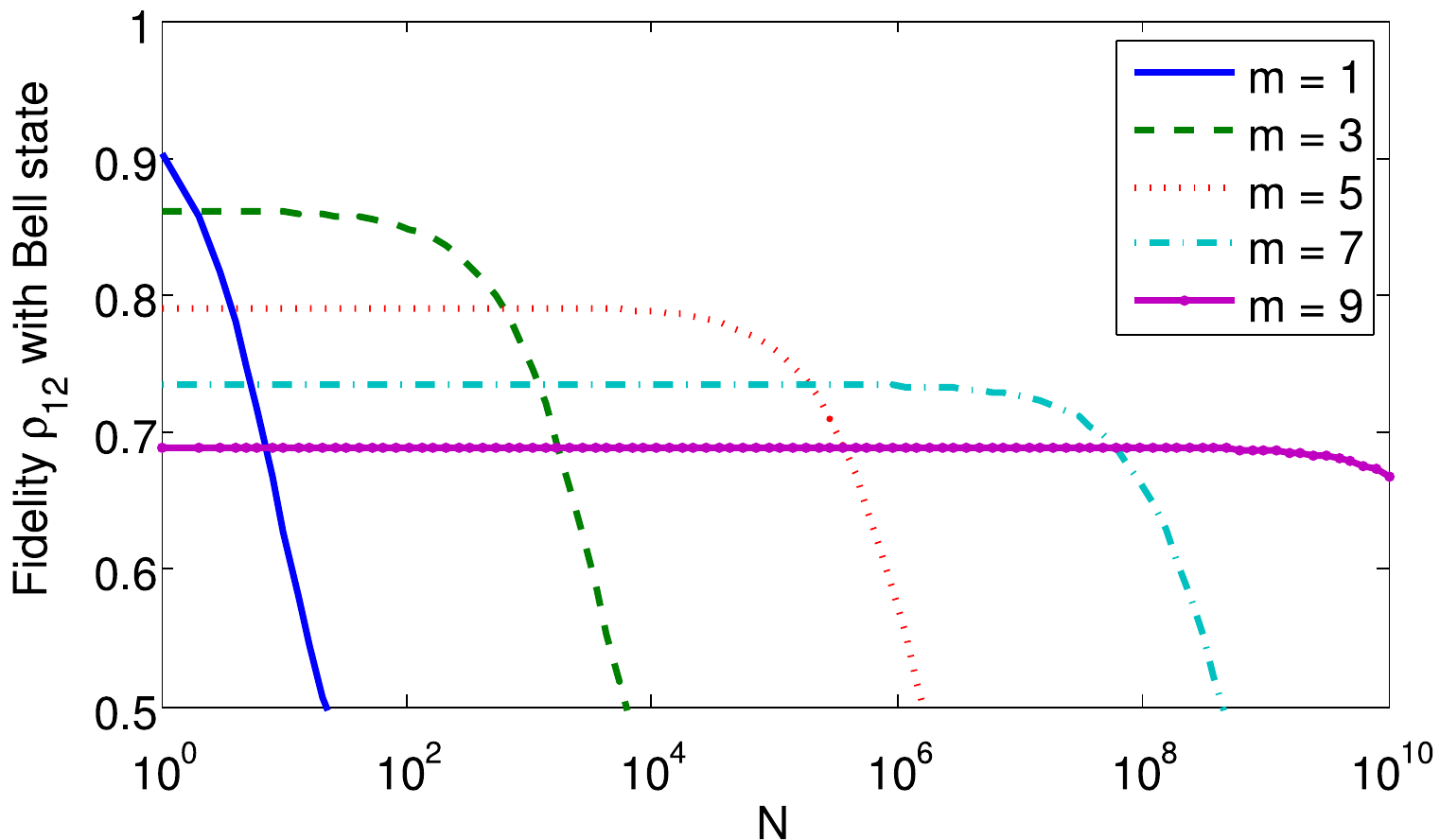}}
\caption[]{\label{fig:dist} (Color online) Example of distillability properties of C-GHZ state. For a fixed decoherence time $p=0.9$, the fidelity $F$ in  Eq.~(\ref{eq:17}) is calculated for various block size $m$ and logical party-number $N$. As long as $F>1/2$, the resource state contains distillable $N$-party entanglement. For $m = 9$ we are able to distill Bell pairs even for $N = 10^{10}$, while for the standard GHZ state ($m = 1$) we stay much below $N = 10^2$.}
\end{figure}

We examine the first two points of the protocol. For simplicity, we always assume the outcome $\lambda_{+}$ for all measurements. After normalization, we obtain the two-particle reduced density operator
\begin{equation}
  \label{eq:16}
  \begin{split}
    \rho_{12} &= \frac{1}{8d^{2}}\left[
      \begin{pmatrix}
        2c_0^{+} &  p^m\\
        p^{m} &  2c_0^{+}\\
      \end{pmatrix}^{\otimes 2} +
      \begin{pmatrix}
        2c_0^{+} &  -p^m\\
        -p^{m} &  2c_0^{+}\\
      \end{pmatrix}^{\otimes 2}\right.\\ & +\left.
      \begin{pmatrix}
        2c_0^{-} &  -p^m\\
        p^{m} &  -2c_0^{-}\\
      \end{pmatrix}^{\otimes 2} +
      \begin{pmatrix}
        2c_0^{-} &  p^m\\
        -p^{m} &  -2c_0^{-}\\
      \end{pmatrix}^{\otimes 2}\right],
  \end{split}
\end{equation}
with $c_0^{\pm}$ from Eq.~(\ref{eq:45}). The fidelity of this density operator with$\ket{\Phi_L^{+}}$ equals
\begin{equation}
  \label{eq:17}
  F = \frac{1}{4}\left\{ 1+\frac{p^{2m}}{4 c_0^{+ 2}} \left[1+\left(\frac{c_0^{-}}{c_0^{+}}\right)^{N-2}\right]+\left(\frac{c_0^{-}}{c_0^{+}}\right)^N \right\}.
\end{equation}
A numerical example for a fixed decoherence time is shown in Fig.~\ref{fig:dist}. From this plot we see that increasing the logical block size $m$ leads to an exponentially fast increasing of the maximal system size with which distillation is possible. Notice that for $p=0.9$, the maximum system size is increased from $N=53$ for $m=1$ (no encoding) to $N \approx 10^{12}$ for $m=10$. The key element for this result is that the only term that depends on $N$ is the ratio $c_0^{-}/c_0^{+}$, which is extremely robust against noise, if we increase $m$.

In fact, one can show, like in Sec.~\ref{sec:results-c-ghz}, that a logarithmic growth of $m$ with the logical system size $N$ always guaranties a finite decoherence time such that, for sufficiently large $N$, we have $F>1/2$. To see this, we set again $N = b^m$, with $b>1$. First we have a look at the term $(c_0^{-}/c_0^{+})^{N} = (c_0^{-}/c_0^{+})^{b^m}$. For large, but finite $m$ we use the short hand notation $y = (1-p)/(1+p)$ and approximate this term by
\begin{equation}
  \label{eq:25}
\left(\frac{1-y^m}{1+y^m}\right)^{b^m}\approx \left(1-2y^m\right)^{b^m} \approx 1- 2\left(by\right)^m.
\end{equation}
The approximation $(c_0^{-}/c_0^{+})^{N}\approx 1- 2\left(by\right)^m$ is valid if $by < 1$, i.e.~for a given time $t$ we can choose a maximal basis $b<1/y$. The same procedure leads to $(c_0^{-}/c_0^{+})^{N-2}\approx 1- \left(2b^{m}-4\right) y^m$. The total fidelity from Eq.~(\ref{eq:17}) then reads approximately $\mathcal{F}\approx 1/2\left[1+f(m)\right]$ with
\begin{equation}
  \label{eq:26}
  f(m) = \frac{p^{2m}}{4c_0^{+ 2}}\left(1+2y^{m}\right) - (by)^m \left(1+\frac{p^{2m}}{4c_0^{+ 2}}\right).
\end{equation}

In the following, we convince ourselves that --for a fixed $b$-- there exists a sufficiently large block size $m$ so that we can find a finite time interval $[0,t_0]$ in which the C-GHZ state is distillable; that is, $\mathcal{F}>1/2$. To show $f(m)>0$, we consider the inequality
\begin{equation}
  \label{eq:27}
  \frac{p^{2m}}{4c_0^{+ 2}}\left(1+2y^{m}\right) > (by)^m \left(1+\frac{p^{2m}}{4c_0^{+ 2}}\right).
\end{equation}
We notice that $c_0^{+} = \left(\frac{1+p}{2}  \right)^{m} \frac{1+y^m}{2}$ and apply again the approximation $y^m \ll 1$ to see that the left-hand-side of inequality (\ref{eq:27}) equals approximately $\left[2p/(1+p)\right]^{2m}$. The right-hand-side is simply approximated by $(by)^m$. This leads to a simplified inequality $\left[2p/(1+p)\right]^{2m}>(by)^m$ which is fulfilled as long as $p>p_0 = (1+4/b)^{-1/2}$. This shows that for larger and larger block sizes $m$ we approach a time period $[0,(2\gamma)^{-1} \ln (1+4/b)]$ in which the C-GHZ state is distillable for any finite $m$. Note that the results are valid for any finite $N$, but are not conclusive if we consider the limit $N \to \infty$ and $m \to \infty$.

\subsection{Negativity as a measure of entanglement}
\label{sec:negat-as-meas}

We now turn to a computable bipartite, entanglement measure, the negativity \cite{negativity}. For a given state $\rho$, the system $\mathcal{H}$ on which $\rho$ is defined is divided into two parts $\mathcal{H} = \mathcal{H}_A\otimes \mathcal{H}_B$. The density operator is partially transposed for one of the subsystems, say $A$. For a certain decomposition $\rho=\sum_k c_k A_k\otimes B_k$, the partial transpose reads $\rho^{T_A} =\sum_k c_k(A_k)^T \otimes B_k$. The negativity $\mathcal{N}$ is then defined as the absolute sum of all negative eigenvalues of $\rho^{T_A}$ or equally \cite{NoteTraceNorm}
  \begin{equation}
    \label{eq:4}
    \mathcal{N} = \frac{\lVert\rho^{T_A}\rVert_{1}-1}{2}.
  \end{equation}
Note that all eigenvalues of $\rho^{T_A}$ are positive if $\rho$ is separable with respect to the partition $A:B$ \cite{sepCriterion}.

In the case of multipartite systems, there are several ways to split up the quantum state. For the GHZ state, the two choices $N/2:N/2$ and $1:N-1$, correspond to extreme cases. Here, we focus on the splitting $1:N-1$, since it is the bipartition which is the most fragile under the influence of noise (i.e., it is the one where the negativity vanishes first). The reason for the increased stability of other splittings $k:N-k$ with $k >1$ for the GHZ states is the following. The GHZ states exhibits only one off-diagonal element, which is moved due to the partial transposition to a different entry in the matrix representation. The eigenvalue of the resulting operator is determined by the two-by-two matrix involving this off-diagonal element, and the corresponding diagonal elements where $k (N-k)$ are the number of ones (zeros) of the corresponding diagonal elements. It turns out that the diagonal elements of the noisy GHZ state become smaller with increasing $k$ (as long as $0 \leq k \leq N/2$), which can be easily understood since errors at $k$ qubits are required to contribute to such a diagonal element. As a negative eigenvalue only arises if the product of the off-diagonal elements is larger than the product of the diagonal elements, smaller diagonal elements are favorable for negative eigenvalues. This finally implies that splittings $k:N-k$ with larger $k$ are more stable, where the $N/2:N/2$ splitting is the most stable one \cite{ConcatEC,simon,aolita}. A similar observation holds for the C-GHZ states, and we hence concentrate on the most fragile bipartition $1:N-1$.

\begin{figure}[htbp]
\centerline{\includegraphics[width = 1\columnwidth]{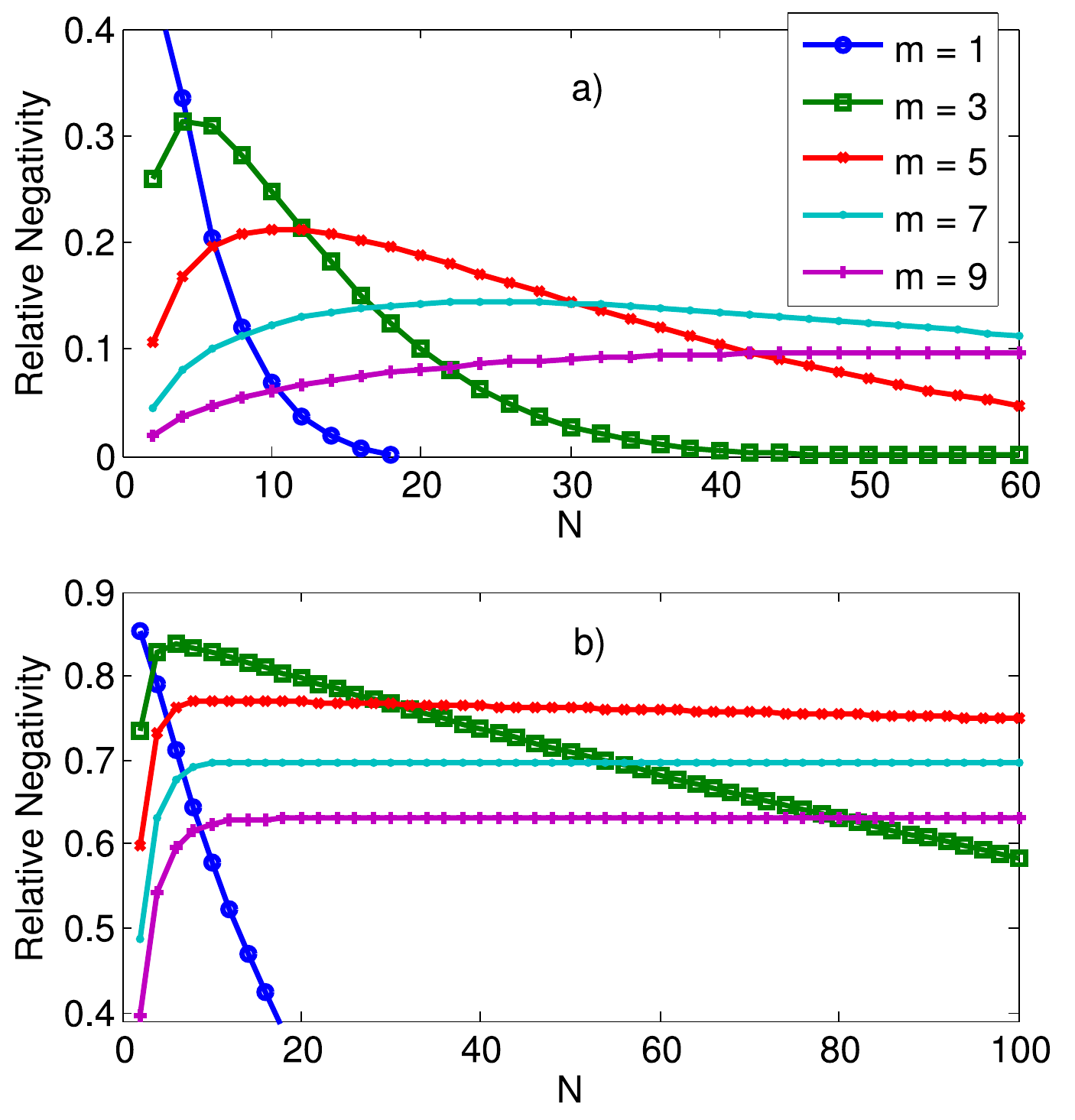}}
\caption[]{\label{fig:Neg} (Color online) Negativity for a fixed noise parameter $p = 0.8$ in (a) and $p = 0.95$ in (b) plotted relative to the noiseless case, which equals 0.5 for all $m,N$. It stabilization effect achieved by increasing the block size $m$ can be seen clearly.}
\end{figure}

In the following, we review the results of Ref.~\cite{cGHZ} concerning the calculation of the negativity for the C-GHZ state Eq.~(\ref{eq:39}). The computation and the interpretation are explained in more detail. We calculate $\mathcal{N}$ for several noise parameters $p$ and system sizes $m,N$. We show exemplarily that increasing the block size $m$ leads to an exponentially fast stabilization of the negativity with respect to the scaling with $N$, see Fig.~\ref{fig:Neg} for $p = 0.8$ and $p = 0.95$ and also Fig.~1)~a in Ref.~\cite{cGHZ} for $p = 0.9$. Before we discuss the quantitative behavior, we first investigate the origin of the encountered stability.

\subsubsection{Where does the stabilization effect come from?}
To answer this question, we have to look closer at the details of the calculation. The first step is to perform a block-local rotation on every logical qubit, which does not change the negativity. This basis change maps $\ket{0_L} \leftrightarrow \ket{0}^{\otimes m}$ and  $\ket{1_L} \leftrightarrow\ket{1}^{\otimes m}$. The diagonal elements $D =  \mathcal{D}(\ketbrad{0_L})^{\otimes N} + \mathcal{D}(\ketbrad{1_L})^{\otimes N}$ become diagonal in the $\{\ket{0},\ket{1}\}$ basis,
  \begin{equation}
    \label{eq:22}
\tilde{D} =\begin{pmatrix}
\tilde{c}^{+}_{0} & & & \\
      & c^{+}_{1} & & \\
     && \ddots & \\
&&  & \tilde{c}^{+}_{m}
   \end{pmatrix}^{\otimes N} +
   \begin{pmatrix}
      \tilde{c}^{+}_{m} & & & \\
      &\ddots & & \\
     &&  c^{+}_{1} & \\
     &&  & \tilde{c}^{+}_{0}
   \end{pmatrix}^{\otimes N}
  \end{equation}
[$c_i^{\pm}$ from Eq.~(\ref{eq:45})] while the transformation maps the interference terms $O =  \mathcal{D}(\ketbra{1_L}{0_L})^{\otimes N} + h.c.$ to
   \begin{equation}
    \label{eq:23}
    \tilde{O} = \begin{pmatrix}
      0& & & & \tilde{c}^{-}_0\\
      & c_1^{-} & & &\\
      & & \ddots & & \\
      & & & c_{m-1}^{-} & \\
      \tilde{c}^{-}_m & & & &0
   \end{pmatrix}^{\otimes N} + h.c.
 \end{equation}
For both equations we use $\tilde{c}^{\pm}_0=c_0^{\pm}+p^m/2$ and $\tilde{c}^{\pm}_m=c_m^{\pm}-p^m/2$. The transformed state now reads $\tilde{\rho} =  \tilde{D}/2 + \tilde{O}/2$.

Next we perform the partial transposition. We agreed on the splitting $1:N-1$, hence we have to interchange $\tilde{c}^{-}_0$ and $\tilde{c}^{-}_m$ in the first matrix of  Eq.~(\ref{eq:23}). All other elements are not affected. The calculation of single eigenvalues in this expression is very simple, since $\rho^{T_A}$ becomes block diagonal with single numbers and two-by-two matrices of the form $ \left(\begin{array}{cc}
a & b \\
b & a \\
\end{array}
\right)$ with eigenvalues $\lambda^\pm = a \pm b$. The latter ones appear whenever the elements $\tilde{c}^{-}_0$ and $\tilde{c}^{-}_m$ are involved. It is clear that we need, on every parties side, at least one such off-diagonal element to hope for a negative eigenvalue. We now divide the eigenvalues into groups $G_i$. The criterion that an eigenvalue belongs to $G_i$ is the number $N-i$ of off-diagonal elements $\tilde{c}^{-}_0$, which were taken for the off-diagonal entries of the corresponding two-by-two matrix [plus hermite conjugate, see Eq.~(\ref{eq:23})]. As an instance we look at the (only) two by two matrix where we exclusively took off-diagonal elements $\tilde{c}^{-}_0$, (i.e., group $G_0$). The resulting matrix reads
 \begin{equation}
   \label{eq:24}\frac{1}{2}
   \begin{pmatrix}
\tilde{c}^{+ (N-1)}_0 \tilde{c}^{+}_m+ \tilde{c}^{+}_0\tilde{c}^{+ (N-1)}_m & \tilde{c}^{- N}_0  + \tilde{c}^{- N}_m \\
\tilde{c}^{- N}_0  + \tilde{c}^{- N}_m      &
\tilde{c}^{+ (N-1)}_0 \tilde{c}^{+}_m+ \tilde{c}^{+}_0\tilde{c}^{+ (N-1)}_m
   \end{pmatrix}.
 \end{equation}

Taking into account more diagonal elements results in other groups of eigenvalues in our categorization. Generally, the eigenvalues take the form $\lambda^\pm_{\mathbf{k}} = a_{\mathbf{k}} \pm b_{\mathbf{k}}$ where $\mathbf{k} = \left( k_0,k_1,\dots,k_m \right) \in \{1,\dots,N-1\}^{\times (m+1)}$ and
\begin{equation}
\label{eq:42}
\begin{split}
a_{\mathbf{k}} & = \left[\tilde{c}_0^{+ k_0} \tilde{c}_m^{+ (k_m+1)} + \tilde{c}_m^{+ k_0} \tilde{c}_0^{+ (k_m+1)} \right]\prod_{j=1}^{m-1}c^{+ k_j},\\
b_{\mathbf{k}} & = \left[\tilde{c}_0^{- (k_0+1)} \tilde{c}_m^{- k_m} + \tilde{c}_m^{- (k_0+1)} \tilde{c}_0^{+ k_m} \right]\prod_{j=1}^{m-1}c^{- k_j},
\end{split}
\end{equation}
with the side condition $S \Def \sum_{j=0}^m k_j= N-1$ (this eigenvalue belongs to the group $G_{N-1-k_0}$). Involving more and more diagonal elements leads to an exponentially (in $m$ and $N$) large number of eigenvalues. Fortunately, we encounter large degeneracies and can compute the sum of the negative eigenvalues more efficiently. The number of eigenvalues that build up by Eq.~(\ref{eq:42})
is
\begin{equation}
\label{eq:41}
d_{\mathbf{k}} = \left( N-1 \right)!\prod_{j=0}^m\frac{\binom{m}{j}^{k_j}}{ k_j! },
\end{equation}
which includes degeneracies within a block [numerator of Eq.~(\ref{eq:41})] and all possible configurations of elements among the $N-1$ blocks.

In total, the negativity of the C-GHZ state can be written as
\begin{equation}
\label{eq:46}
\mathcal{N} = \frac{1}{2}\sum_{\mathbf{k}:S= N-1}d_{\mathbf{k}}\left( \left| \lambda_{\mathbf{k}}^{-} \right| - \lambda_{\mathbf{k}}^{-}\right)
\end{equation}

\begin{figure}[htbp]
\centerline{\includegraphics[width=1\columnwidth]{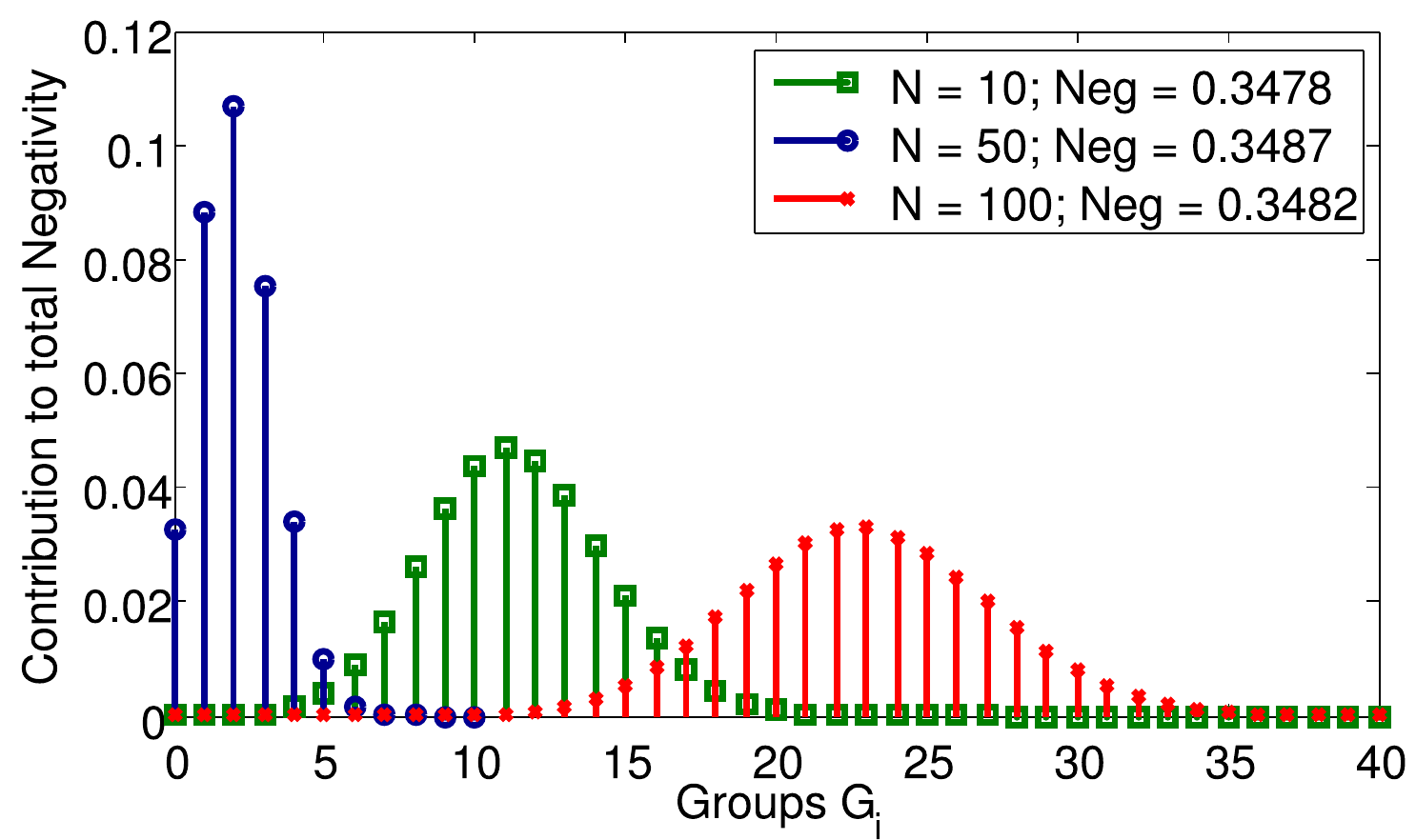}}
\caption[]{\label{fig:gaussian} (Color online) For fixed values of $m$ and $p$ (here $m = 7$ and $p = 0.95$), the contributions from the single groups $G_i$ are plotted (see text). Notice that multiplicative factors are taken into account. The larger the system size $N$, the more important are groups with a relatively large number of diagonal elements for the single two-by-two matrices. The total negativity for these examples is given in the legend.}
\end{figure}

For large system sizes, the total number of possible negative eigenvalues is too large to calculate them all. The above mentioned grouping into sets of eigenvalues therefore serves as a cutoff criterion, which we use to establish a lower bound on the negativity by considering only a subset of the negative eigenvalues. We start with the eigenvalues of matrix (\ref{eq:24}), the only elements of $G_0$; next we compute the contribution to $\mathcal{N}$ for all eigenvalues with one diagonal element $G_1$, continue with $G_2$ and so on. For sufficiently large systems we notice that these contributions follow a Poissonian or Gaussian law (see Fig.~\ref{fig:gaussian}). If we overcome the hump of the Gaussian and the contributions decay exponentially fast, we stop the calculation. That means that we neglect the contributions of groups $G_i$, if it drops below a number between $10^{-6}$ and $10^{-10}$. In this way, we obtain accurate lower bounds on the negativity, which are shown in the plots presented here.

Besides this practical aspect, we can also learn something about the stabilization effect for the negativity. For large system sizes, the entanglement is now constituted from a large number of small negative eigenvalues compared to the standard GHZ state, where we have for all $N$ only one negative eigenvalue.

\subsubsection{The influence of $m$ on the decay of $\mathcal{N}$}
\begin{figure}[htbp]
\centerline{\includegraphics[width = \columnwidth]{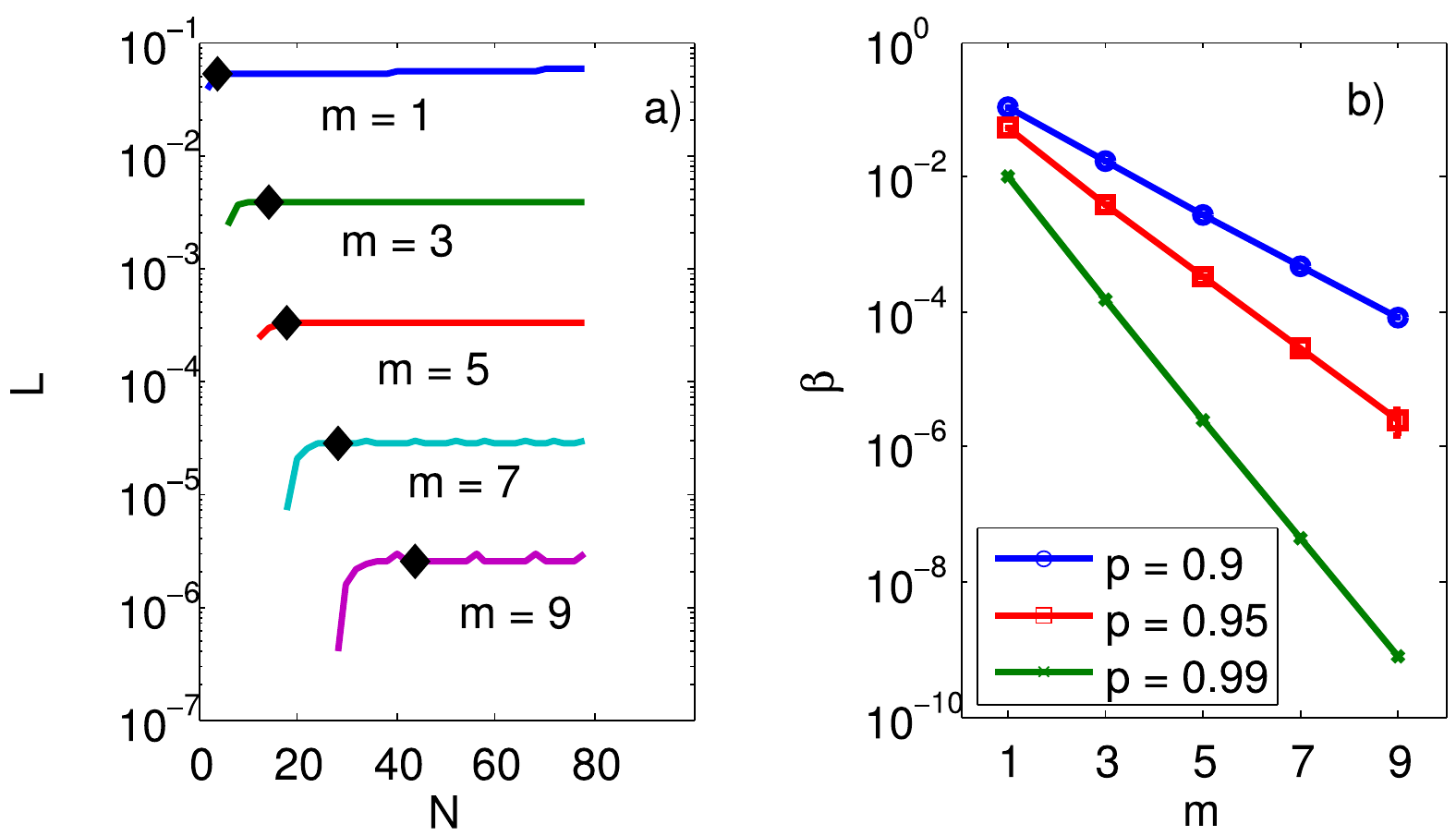}}
\caption[]{\label{fig:expdecay} (Color online) (a) The negative derivative of the logarithm $L$ of $\mathcal{N}$ is calculated and plotted for the example $p = 0.95$ and $m = 1,3,\dots,9$. We can clearly identify constant regions which is equivalent to an exponential decay of $\mathcal{N}$ with respect to $N$. The black diamonds indicate the values taken for the right figure. (b) For three different values of $p$, we calculate the decay rate $\beta$ by identifying the constant regions of $L$ like in part (a) of this figure. It can be clearly seen that $\beta$ decays exponentially fast with $m$, which indicates a stabilization effect.}
\end{figure}

For small interaction times (i.e., $p\gtrapprox 0.9$), we find regions of $N$ where $\mathcal{N}$ can be well approximated by an exponential function. We now investigate how the exponential behaves by an increment of the block size $m$. We find that the decay in turn drops down exponentially fast with $m$. To show this, we first compute the negative derivative of the logarithm of the negativity $L = -\frac{d}{dN} \ln \mathcal{N}$ numerically. For an exponential function $a \exp(-\beta N)$, we would get $L = \beta = \mathrm{const}$. If we numerically encounter intervals for which the actual $L$ is roughly constant, we conclude that, for these values $N$, we can approximate $\mathcal{N}$ by a function $a \exp(-\beta N)$. In Fig.~\ref{fig:expdecay} (a) we indeed find such regions. We repeat this procedure for other noise parameter $p$ and plot the result in Fig.~\ref{fig:expdecay} (b). The coefficient $\beta$ decays for all computed $p$ exponentially fast with $m$. We therefore conclude that, for this choice of bipartition, the negativity can be stabilized exponentially fast by increasing the block size $m$.

We notice that we are currently not able to predict the size of the interval for which an approximation by an exponential is justified. Therefore, it is not guaranteed that an extrapolation to larger system sizes leads to accurate results.

\subsubsection{Negativity as a function of decoherence time}
\begin{figure}[htbp]
\centerline{\includegraphics[width = \columnwidth]{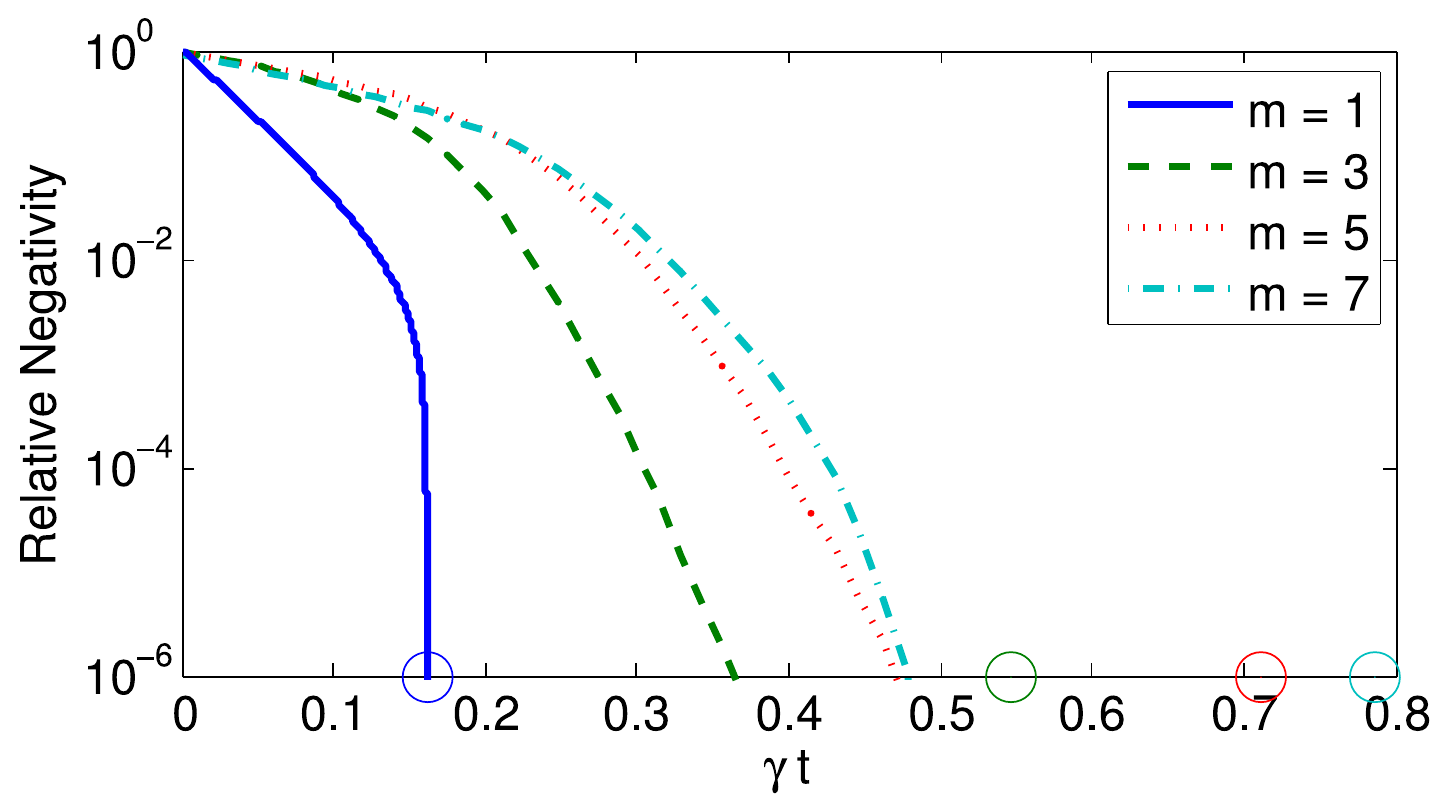}}
\caption[]{\label{fig:negVarP} (Color online) Negativity is calculated relative to  noiseless case $t = 0$ for fixed system size, here $N = 30$. We see that increasing $m$ leads to a longer lifetime of bipartite entanglement. Note that, for this plot, we calculated the contributions from all groups $G_i$. The circles on the time axis give a lower bound on the lifetime of negativity from the calculation of a single negative eigenvalue (see text).}
\end{figure}

So far, we have discussed the behavior of the negativity $\mathcal{N}$ for a fixed $p$ (i.e., a fixed interaction time with the environment). We shortly note that we also see a stabilization effect of $\mathcal{N}$, if we fix the number of logical blocks $N$ and look at the decay of $\mathcal{N}$ as a function of time. In Fig.~\ref{fig:negVarP} we exemplarily demonstrate the improvement of entanglement lifetime by a larger block size $m$. In contrast to a variation with $N$, there is no exponential gain.

As in the case of varying $N$, similar plots like Fig.~\ref{fig:gaussian} can be considered for different values of $p$. Then one sees that the smaller $p$, the more important are groups $G_i$ with large $i$.

\begin{figure}[htbp]
\centerline{\includegraphics[width=\columnwidth]{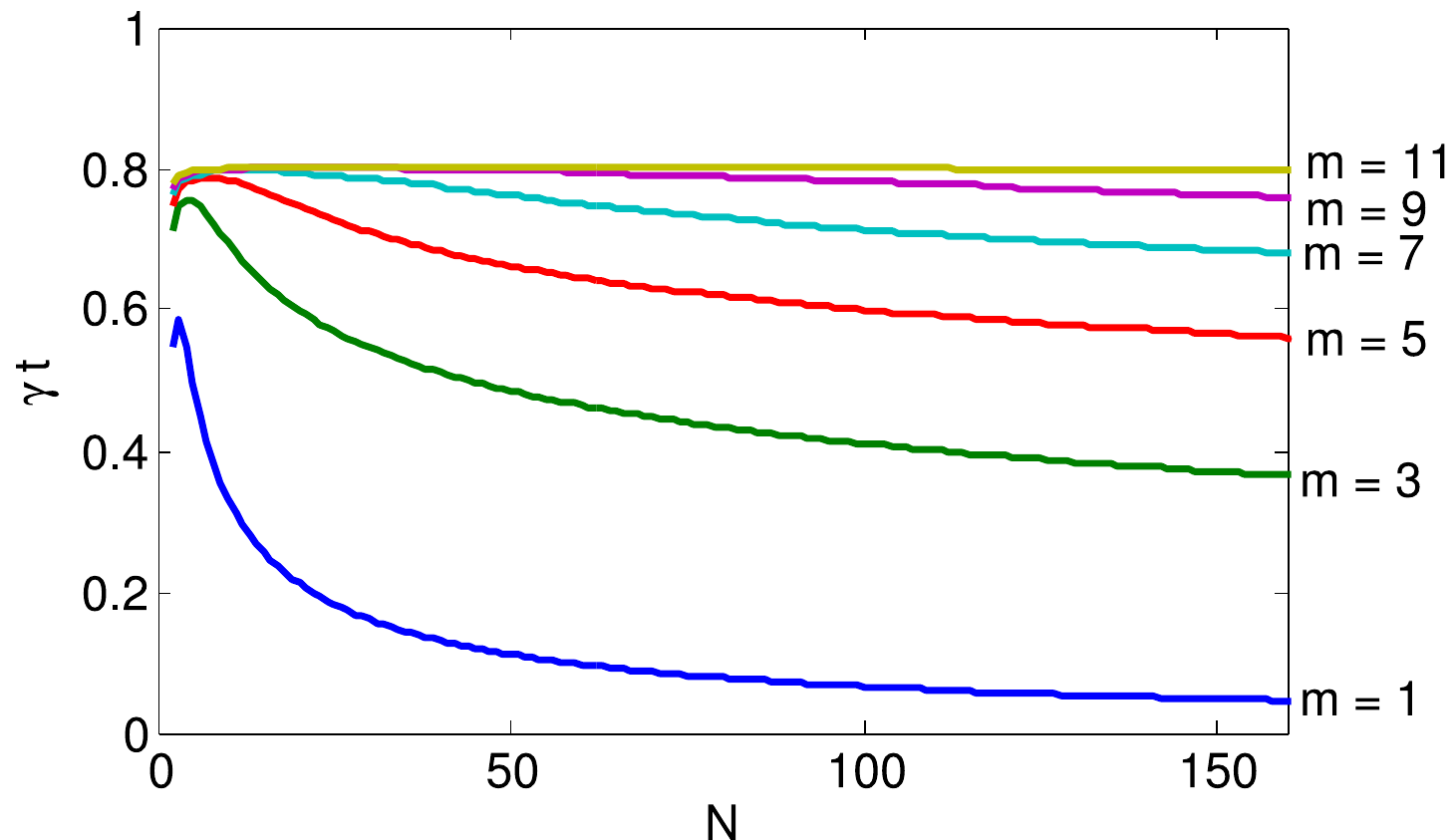}}
\caption[]{\label{fig:lifetime} (Color online) Lower bound on lifetime of negativity plotted for various system parameters $m,N$.}
\end{figure}

To obtain a lower bound on the lifetime of the negativity (i.e., the time until which the negativity remains nonzero), it is enough to study a single negative eigenvalue. We expect the negative eigenvalue of matrix (\ref{eq:24}) to survive longest. We investigate for which time the smaller eigenvalue of matrix (\ref{eq:24}) stays negative. This time is then a lower bound on the entanglement lifetime. The (numerical) result is shown in Fig.~\ref{fig:lifetime}. We observe that the maximum lifetime cannot be arbitrarily extended by an increase of $m$ and is approximately bounded $\gamma t \gtrapprox 0.80$. For larger $N$ we need a larger $m$ to come close to this bound.

\subsubsection{Upper bound on the negativity for general logical GHZ states}
The C-GHZ state (\ref{eq:2}) is formed by two states that are fully separable with respect to the block structure. For general states of this kind we are able to give an upper bound for the negativity (\ref{eq:4}) in terms of the trace norm of the interference terms, see Sec. \ref{sec:diagonal-elements}. Therefore, we consider the following class of states
  \begin{equation}
    \label{eq:18}
    \ket{\psi} = \frac{1}{\sqrt{2}}\left(\ket{\psi_0} + \ket{\psi_1} \right),
  \end{equation}
  with $\ket{\psi_0} = \ket{\psi_0^A}\otimes \ket{\psi_0^B}$ and $\ket{\psi_1} = \ket{\psi_1^A}\otimes \ket{\psi_1^B}$; that is, both constituting states are separable with respect to a certain splitting $A:B$. We now consider this state vector (\ref{eq:18}) under the influence of some uncorrelated noise $\rho = \mathcal{E}_t(\ketbrad{\psi})$ and calculate the negativity regarding this bipartition. First we use the triangular inequality
  \begin{equation}
    \label{eq:20}
    \begin{split}
      2\lVert\rho^{T_A}\rVert_{1} &\leq \lVert
      \mathcal{E}_t(\ketbrad{\psi_0})^{T_A} \rVert_{1} + \lVert
      \mathcal{E}_t(\ketbrad{\psi_1})^{T_A}\rVert_{1} \\ &+
      \lVert\mathcal{E}_t(\ketbra{\psi_0}{\psi_1})^{T_A}\rVert_{1} +
      \lVert\mathcal{E}_t(\ketbra{\psi_1}{\psi_0})^{T_A}\rVert_{1}.
    \end{split}
  \end{equation}

The first two terms on the rhs of Eq.~(\ref{eq:20}) are unity since neither the partial transpose with respect to party $A$ nor the influence of the noise affect the trace norm. The third and fourth term are invariant under partial transpose because we assumed separability for $\ket{\psi_0}$ and $\ket{\psi_1}$ for the given bipartition and the noise process affects the parts individually. Using the definition of negativity [Eq.~(\ref{eq:4})], we end up with the upper bound
  \begin{equation}
    \label{eq:21}
    \mathcal{N} \leq \lVert\mathcal{E}_t(\ketbra{\psi_0}{\psi_1})\rVert_{1}/2
  \end{equation}
We have therefore shown that the negativity for special states like the logical GHZ state Eq.~(\ref{eq:2a}) with any encoding can be estimated from above by the trace norm of the noisy interference terms. Recall that in this case the trace norm decays always exponentially with the number of blocks $N$.

Even though the derived bound is not sharp for the C-GHZ state, it nicely reflects the qualitative scaling behavior in both system size $N$ and block size $m$. That is, both the upper bound as well as the actual results for negativity of noisy C-GHZ states show an exponential decay with system size $N$, where the rate is (exponentially) reduced with increasing block size $m$.

  \subsection{Genuine multipartite entanglement}
  \label{sec:genu-mult-entangl}

We now turn to another aspect of multiparticle entanglement; namely, genuine multipartite entanglement (GME) \cite{Otfried,GME}. Quantum states containing GME are not decomposable into an incoherent sum of bipartite-entangled states. The noiseless GHZ state shows GME, but --without surprise-- also here the entanglement lifetime vanishes exponentially fast with the system size in the presence of decoherence.

  For states that are diagonal in the GHZ basis, in Ref.~\cite{Otfried}  a necessary and sufficient criterion for the appearance of GME was shown. A GHZ-diagonal state is genuine multipartite entangled, if and only if one of its coefficients in this basis is larger than $1/2$.

The C-GHZ state can be brought into a GHZ-diagonal state by locally transforming every block. In particular, we apply a local unitary operation that maps the standard $\ket{i}$-basis ($i = 0,\dots,2^m-1$) to the GHZ basis $1/\sqrt{2}\left( \mathbbm{1}^{\otimes m}+ \sigma_x^{\otimes m} \right)\ket{i}$ [for $i = 0,\dots, (2^m-1)/2$; $\sigma_x^{\otimes m}\ket{i}$ is mapped to $1/\sqrt{2}\left( \mathbbm{1}^{\otimes m}- \sigma_x^{\otimes m} \right)\ket{i}$]. Now we are in the position to calculate the coefficients in the GHZ-basis of the total space $\mathbbm{C}^{2\otimes mN}$. The largest coefficient $\alpha$ is easily found and equals
  \begin{equation}
\label{eq:36}
\begin{split}
  \alpha &= \frac{1}{2}\left[\left( c_0^{+}+p^m/2 \right)^N + \left( c_0^{+}-p^m/2    \right)^N\right. \\ & + \left.\left( c_0^{-}+p^m/2 \right)^N + \left( c_0^{-}-p^m/2 \right)^N \right].
\end{split}
\end{equation}
We quickly see that $\alpha$ is unity for $p=1$ (noiseless) and drops down exponentially fast with $N$ and $m$. In contrast to other properties so far, the result is that the C-GHZ state cannot maintain its GME for larger times by increasing the block size $m$ but it looses this kind of multipartite entanglement even faster with larger $m$.

One may also allow for stochastic local operations, in particular projections to certain subspaces, and investigate whether the resulting states are GME. This would indicate that GME is already present in the initial state. We therefore project every logical block into the subspace spanned by $\ket{0}^{\otimes m}$ and $\ket{1}^{\otimes m}$. Note that the projection does not destroy entanglement, since there are no correlations between the projected subspace and the rest. The total state can be effectively described as a $N$-qubit state. Again, we identify the largest overlap $\tilde{\alpha}_0$ of this state with a GHZ state and find
\begin{equation}
\label{eq:37}
 \begin{split}
\tilde{\alpha}_0  &= \frac{1}{2^{N+1}}\left[\left( 1+\frac{p^m}{4c_0^{+}} \right)^N +\left( 1-\frac{p^m}{4c_0^{+}} \right)^N \right. \\ & +\left. \left( \frac{c_0^{-}}{c_0^{+}}+\frac{p^m}{4c_0^{+}} \right)^N +\left( \frac{c_0^{-}}{c_0^{+}}-\frac{p^m}{4c_0^{+}} \right)^N \right].
\end{split}
\end{equation}
This new expression improves the situation for $m>1$ and indeed the GME-lifetime for the C-GHZ state for $m=2$ is larger than for $m=1$ (standard GHZ). However, further increasing $m$ again leads to shorter and shorter lifetimes. The projection into other subspaces does not lead to any state that is genuine multipartite entangled.

We therefore have to conclude that the GHZ encoding (\ref{eq:1}) does not stabilize GME but leads to shorter lifetimes. An interesting point in this context is the observation that the Fisher information for the C-GHZ state (see Ref.~\cite{cGHZ}) can indeed be stabilized in contrast to GME. The Fisher information is an indicator for the usefulness of a given mixed state for parameter estimation with improved sensitivity beyond the standard quantum limit. The Fisher information for noisy C-GHZ states can be calculated using similar techniques as used for the calculation of the negativity (see Ref.~\cite{cGHZ}), and one observes a stabilization with increasing block size $m$. This shows that for the usability of a given quantum state for parameter estimation, the presence of GME is not necessary.

  \section{Is the C-GHZ state macroscopic under noise?}
  \label{sec:c-ghz-state}

\subsection{Definition of ``macroscopicity'': index $q$}
\label{sec:defint-macr}

What does it mean if we call a quantum state macroscopic? First of all, it should consist of a macroscopic number of ``elementary'' particles. Leggett \cite{Leg80} had a stricter definition in mind. Among others, he analyzed the cat paradox of Schr\"odinger \cite{Schroedinger35} and concluded that a macroscopic quantum state should in principle be able to demonstrate the validity of quantum mechanics on a certain scale. The goal is the realization of an experiment that can neither be explained classically nor by an accumulative microscopic quantum effect.

There have been already several attempts \cite{DSC,p-Index,BM,KWDC,MvD,LJ,qIndex} to classify macroscopic quantum states. Except for Refs.~\cite{LJ,qIndex}, these were formulated only for pure states. Here we focus on one of the exceptions, the so-called index $q$ \cite{qIndex,qInd2}. The intuition of the authors is to identify states that behave differently than classical states. They focus on sums of \textit{local} observables $A = \sum_iA^{(i)}$, since these are typical for large systems (e.g., the magnetization in solids). Given a quantum state $\rho$, we consider $C(\rho) = [A,[A,\rho]]$ like in \cite{qInd2} and maximize its trace norm over all local observables $A$
  \begin{equation}
  \label{eq:19}
  c = \max \left[ N, \max_{A: local}\lVert C(\rho)\rVert_{1}\right].
\end{equation}
We are interested in the order of $c=O(N^q)$ \cite{NoteOrder}. A state $\rho$ is called macroscopic if and only if $q =2$, assuming that the maximal eigenvalue of $A$ equals $N$. For pure states, it can be shown \cite{qIndex}, that $q = 2$ if and only if the variance of $A$ under $\rho$ is in the order of $N^2$. Since uncorrelated states can only exhibit variances $V(A)$ in the order of $N$, a pure state showing $V(A) = O(N^2)$ is definitely nonclassical. For practical purposes, we skip here the maximization over all local observables and use an $A$ which gives a high $c$.

The standard GHZ state is macroscopic due to this criterion, if we choose $A = \sum_{i=1}^N\sigma_z^{(i)}$. Since $\{A,\ketbrad{\mathrm{GHZ}}\} = 0$, i.e.~the projector onto the GHZ state anticommutes with $A$, we end up with $c=4N^2$ in this case, which is the maximal value. The index $q$ has to be extended in order to recognize the C-GHZ state as macroscopic. The term ``local'' now means that every $A^{(i)}$ can act on the whole logical block, i.e.~the superscript means that $A^{(i)}$ acts nontrivially on the block $i\in \{1,\dots,N\}$. For the encoding of  Eq.~(\ref{eq:1}), the choice $A^{(i)} = \ketbrad{0_L}^{(i)} - \ketbrad{1_L}^{(i)}$ leads to the same index $q$ as the standard GHZ state. We remark that a choice $A^{(i)} = \sigma_x^{\otimes m}$ would also be possible here.

\subsection{Numerical results under white noise}
\label{sec:numer-results-under}

In the original attempt of \cite{qIndex,qInd2}, the index $q$ is more qualifying than quantifying the macroscopic property of a state. Under the influence of noise, $c$ changes for the GHZ state (we fix $A$ for all times). For the choice of white noise as decoherence process, we obtain $c = 4 N^2 p^N$.
In the following we will use a renormalized index $q$:
\be
N_{\mathrm{eff}} = c/(4N).
\ee
This may be interpreted as an ``effective size'' of the system, which is supported by comparing the values for pure state with other measures of effective size for macroscopic quantum superpositions (see, e.g., Refs.~\cite{DSC,MvD}). For GHZ states, we obtain a value $N$ as desired, while, for example, for states of the form $|0\rangle^{\otimes N} + |\varphi\rangle^{\otimes N}$ with $|\langle 0 |\varphi\rangle|^2 = 1 - \epsilon^2$  we find $N_{\rm eff}= \epsilon^2 N$, consistent with \cite{DSC}. Whether an interpretation as effective size is also valid for the mixed state case is however not clear.

\begin{figure}[htbp]
\centerline{\includegraphics[width = 1\columnwidth]{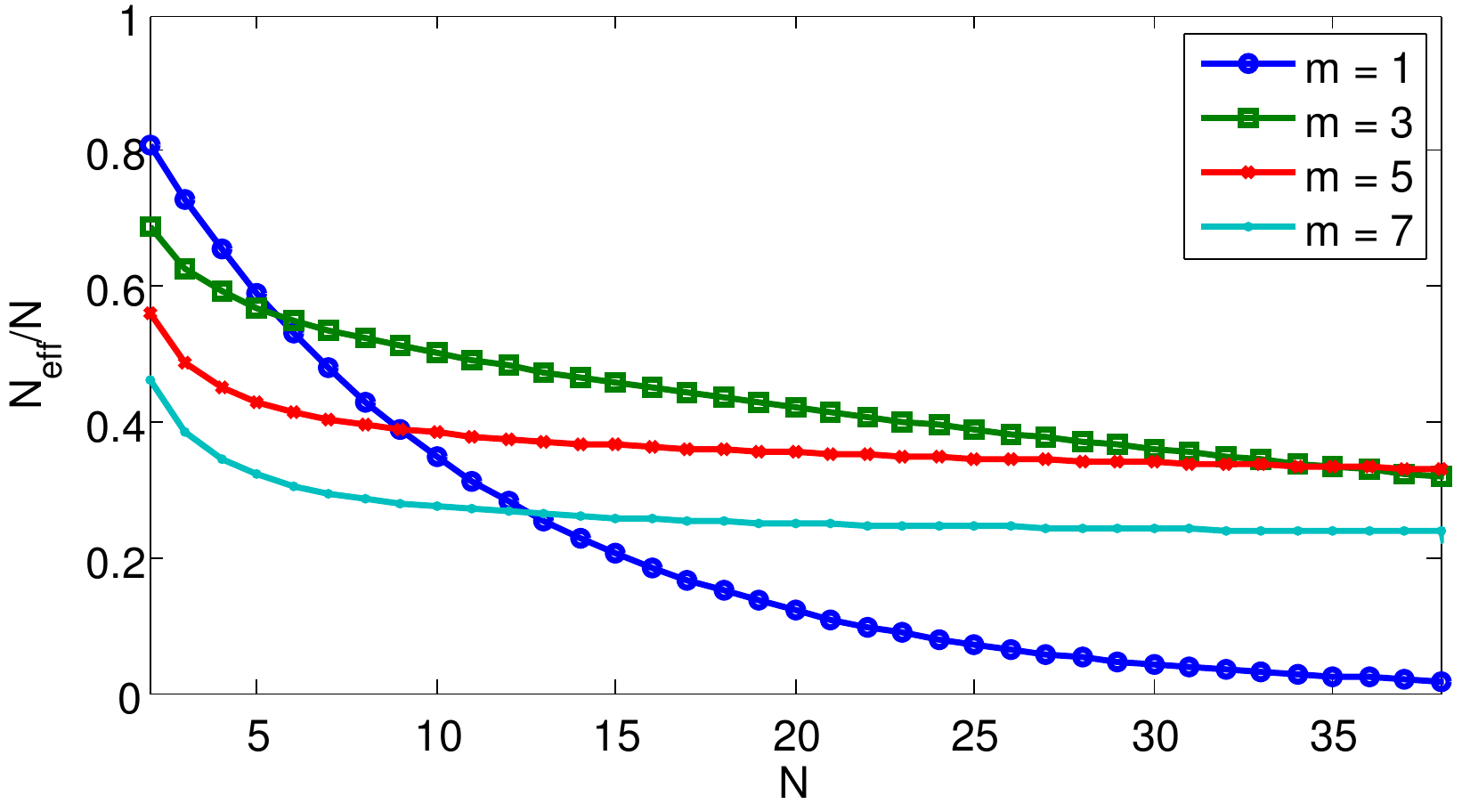}}
\caption[]{\label{fig:qind} (Color online) Calculation of renormalized index $q$ of C-GHZ state. The values are plotted relative to the noise-free case (i.e., $N_{\mathrm{eff}}/N$), for a fixed white noise parameter $p = 0.9$ and for various values of $N$ and $m$. The decay of the relative index $q$ is slowed by increasing the block size $m$.}
\end{figure}

With this assignment, we study the renormalized index $q$ $N_{\rm eff}$ for the C-GHZ state under the influence of white noise decoherence. We have seen that the standard GHZ state exhibits $N_{\mathrm{eff}} = p^{N}N$, which means that $N_{\rm eff}$ decays exponentially fast in time and system size $N$. To compute $N_{\mathrm{eff}}$ for the C-GHZ state, we numerically calculate $c$ for the special choice of $A^{(i)} = \ketbrad{0_L}^{(i)} - \ketbrad{1_L}^{(i)}$. The remaining calculation can be done in a similar manner as in the case of the negativity: First we rotate the state locally to obtain the Eqs.~(\ref{eq:22}) and (\ref{eq:23}). From this we calculate the action of $A$ entry-wise. Like the computation, the results are qualitatively similar to the negativity. We observe that increasing the number $m$ of physical qubits forming a logical block leads to a stabilization of $N_{\rm eff}$ (see Fig.~\ref{fig:qind}). Although we are not able to predict the behavior for system sizes that deserve the name macroscopic, we are confident that using the GHZ-encoding leads to a renormalized index $q$ which is more resistant in a noisy environment. Therefore, the C-GHZ state of a large system size $N$ could stay macroscopic for a longer time period compared to the standard GHZ state if $m$ is large enough.

\subsection{Upper bound for the index $q$}
\label{sec:upper-bound-q}

We now derive an inequality for the index $q$, which shows once again the importance of the off-diagonal elements discussed in Sec.~\ref{sec:diagonal-elements}. We consider therefore the quantum state $\ket{\psi} = \frac{1}{\sqrt{2}}\left(\ket{\psi_{0}}+\ket{\psi_{1}}\right)$ which consists of two states that both obey a ``small'' index $q$ [i.e., behaves like $c=o(N^{2})$ \cite{NoteOrder}]. For this criterion, $\ket{\psi_{0}}$ and $\ket{\psi_{1}}$ are nonmacroscopic. On the other hand, the superposition $\ket{\psi}$ exhibits a macroscopic index $q=2$. The state $\left| \psi \right\rangle$ is now subject to white noise decoherence $\mathcal{D}$ [Eq.~(\ref{eq:5})]. For convenience, we skip the index $t$ in this section. Assuming we have found the optimal $A$ and $c\geq N$, we use the triangle inequality
\begin{equation}
  \label{eq:32}
  \begin{split}
   c  &= \lVert C(\rho)\rVert_1\leq \\ &\frac{1}{2}
\left\{\lVert C\left[\mathcal{D}(\ketbrad{\psi_0})\right]\rVert_{1} +\lVert C\left[\mathcal{D}(\ketbrad{\psi_1})\right]\rVert_{1}\right\} \\
&+ \lVert C\left[\mathcal{D}(\ketbra{\psi_1}{\psi_0})\right]\rVert_{1}.  \end{split}
\end{equation}
Per definition, the index $q$ of $\ket{\psi_0}$ and $\ket{\psi_1}$ are of the order $o(N^2)$ \cite{NoteOrder}. Hence the index $q$ of the first two terms in inequality (\ref{eq:32}) cannot contribute to the $O(N^2)$ term of $c$ \cite{NoteAssumption}. So we are left to investigate $\lVert C\left[\mathcal{D}(\ketbra{\psi_1}{\psi_0})\right]\rVert_{1}$. For all linear operators $\sigma$ the triangle inequality
\begin{equation}
  \label{eq:33}
  \lVert C(\sigma)\rVert_{1} \leq \lVert A^2\sigma \rVert_{1} + \lVert \sigma A^2 \rVert_{1} + 2\lVert A\sigma A \rVert_{1},
\end{equation}
is valid and we use the H\"older inequality \cite{holder} to estimate $ \lVert A^2\sigma \rVert_{1} \leq \lambda_{\mathrm{max}}(A)^2 \lVert \sigma \rVert_{1}$, where $\lambda_{\mathrm{max}}(A) = N$ is the maximal eigenvalue of $A$. The other terms can be estimated in the same way. Hence the main result is
\begin{equation}
  \label{eq:34}
  c \leq 4 N^2 \lVert\mathcal{D}(\ketbra{\psi_1}{\psi_0})\rVert_{1}  + o(N^2).
\end{equation}
This means that the effective size of the state $\ket{\psi}$ can be bounded from above by $\lVert\mathcal{D}(\ketbra{\psi_1}{\psi_0})\rVert_{1}$, the decay of the off-diagonal elements under a decoherence map. Although a large off-diagonal element does not guarantee a high $c$ value, in this scenario it is absolutely necessary.

Again, the derived bound on the index $q$ is not sharp for the C-GHZ state, but shows a good qualitative agreement in scaling with respect to system size $N$ and block size $m$.

\section{Alternative encoding of a logical block}
\label{sec:alternatives}

The C-GHZ state shows impressive stability under uncorrelated decoherence processes. Nevertheless one could ask whether the choice of  Eq.~(\ref{eq:1}) is optimal and whether other pairs of orthogonal, multipartite states could possibly perform similarly. In this section, we give partial answers to this question. First we investigate another important family of quantum states as a basis for logical encoding, the so-called cluster states. Next we are interested in the iteration of the encoding (\ref{eq:1}). Later on, we tackle this questions by a numerical search of the optimal encoding and discuss relations to active quantum error correction. 

In this section, we exclusively concentrate on the decay of the off-diagonal elements under the influence of white noise, see Sec.~\ref{sec:diagonal-elements}. This indicates the usability of the respective encoding since the trace norm gives us an upper bound for other properties (Sec.~\ref{sec:negat-as-meas} and \ref{sec:upper-bound-q}). It is, however, clear that the norm of the interference terms is not enough to judge any other quantum property rigorously; see in this context the discussions at the end of Sec.~\ref{sec:cluster-ghz} and in Sec.~\ref{sec:optim-relat-active}.

The summary of this section is that all attempts to find other codewords for the logical encoding tried so far lead to a norm of the interference terms that is more unstable than that for the GHZ encoding (\ref{eq:1}).

\subsection{Using cluster states to encode quantum information}
\label{sec:using-cluster-states}

Among other important classes of multipartite quantum states, the graph states and in particular the cluster states have been studied intensively \cite{hein06}. Graph states are generated by applying phase gates $C$ between pairs of qubits prepared in a suitable product state. If we impose a certain geometry [e.g., a one-dimensional chain or a two-dimensional (2D) lattice], and apply the phase gates only between nearest neighbors, these states are called cluster states. Most prominent is the 2D cluster state since it can serve as a resource for universal measurement-based quantum computation \cite{ClusterState,RauBroBri03}. Another application is to use 1D cluster states (or other graph states) as codewords for quantum error correction \cite{Grassl02,GraphErrorCorr}. This idea has been iterated in Ref.~\cite{ConcatEC}, which is discussed in Sec.~\ref{sec:conc-encod}.

Here, we are interested in the performance of the logical GHZ state (\ref{eq:2a}) if we exchange the encoding (\ref{eq:1}), by choosing $|\tilde{0}_L\rangle =  \ket{\mathrm{Cl}^{+}_m}$, $|\tilde{1}_L\rangle =  \ket{\mathrm{Cl}^{-}_m}$ with
\begin{equation}
  \label{eq:1c}
  \begin{split}
    \ket{\mathrm{Cl}^{+}_m} &= \prod_{i=1}^m C^{(i,i+1)} \ket{+}^{\otimes m} \\
    \ket{\mathrm{Cl}^{-}_m} &= \prod_{i=1}^m C^{(i,i+1)} \ket{-}^{\otimes m}.
  \end{split}
\end{equation}
The orthonormal basis $\{\ket{+},\ket{-}\}$ is the eigenbasis of the Pauli operator $\sigma_x$. The phase gate reads $C^{(i,i+1)} = \ketbrad{0}^{(i)}\otimes\mathbbm{1}^{(i+1)} + \ketbrad{1}^{(i)}\otimes\sigma_z^{(i+1)}$ and is applied to every two neighbors respecting periodic boundary conditions (i.e., $m+1\equiv 1$).

\begin{figure}[htbp]
\centerline{\includegraphics[width = \columnwidth]{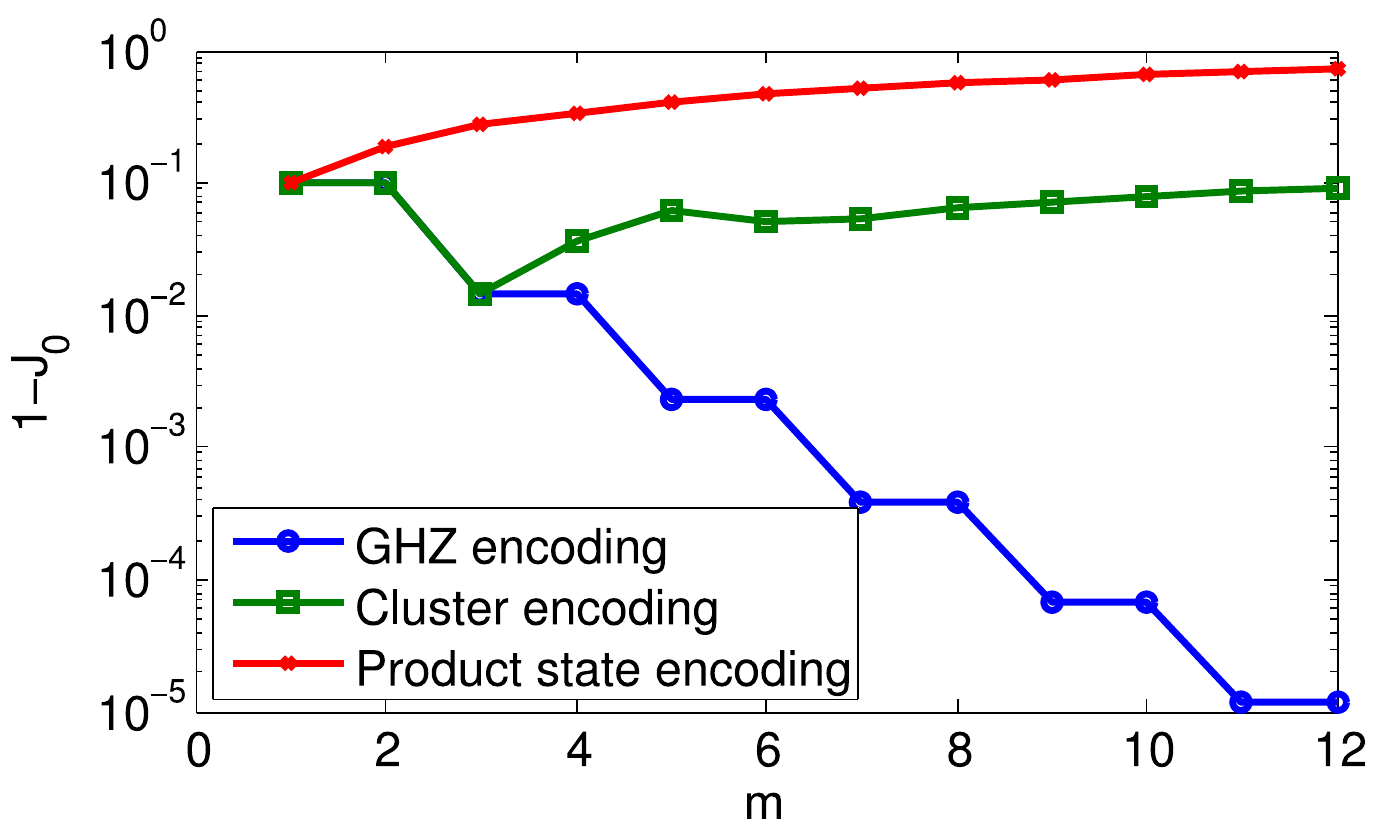}}
\caption[]{\label{fig:ClusterCod1} (Color online) Comparison of different encodings for a logical block of size $m$. The off-diagonal element $\ketbra{\tilde{0}_L}{\tilde{1}_L}$ is subject to white noise and the trace norm $J_0 = \lVert \mathcal{D}_t\left(\ketbra{\tilde{0}_L}{\tilde{1}_L}\right)\rVert_1$ is calculated for a fixed noise parameter, here $p=0.9$, and several block sizes $m$. We see that the cluster encoding [Eq.~(\ref{eq:1c})] for $m$ up to three coincides with the GHZ encoding [Eq.~(\ref{eq:1})], while later it scales similarly like the product state encoding, since it decreases by enlarging $m$.}
\end{figure}

We again consider a single block and calculate the trace norm of $J_{0,\mathrm{Cl}} = \lVert \mathcal{D}_t\left(\ketbra{\tilde{0}_L}{\tilde{1}_L}\right)\rVert_1 = \lVert \mathcal{D}_t\left(\ketbra{\mathrm{Cl^{+}_m}}{\mathrm{Cl^{-}_m}}\right)\rVert_1$ for several block sizes $m$. For small $m$, it is straightforward to obtain an analytical expression; for example, for $m = 5$ we find that $J_{0,\mathrm{Cl}} = \frac{1}{2} p^3 (5 - 3 p^2)$. We see that --similar to the GHZ encoding (\ref{eq:1})-- the first derivative vanishes at $p=1$, while in contrast the second derivative does not. It seems that the second derivative becomes in fact larger and larger with increasing $m$ at $p=1$. This indicates a reduced stability of the interference terms using cluster states for the encoding. To support this numerically, we calculate $J_{0,\mathrm{Cl}}$ for fixed $p<1$ and $m$ up to twelve, see Fig.~\ref{fig:ClusterCod1}. As a reference, we also plot the trace norm of the noisy off-diagonal element $\lVert \mathcal{D}_t\left(\ketbra{0}{1}^{\otimes m}\right)\rVert_1 = p^m$. This encoding simply consists of two product states $\left| \tilde{0}_L \right\rangle = \left| 0 \right\rangle^{\otimes m}, \left| \tilde{1}_L \right\rangle = \left| 1 \right\rangle^{\otimes m}$, which are orthogonal in every qubit and represents somehow the ``worst case''. Remarkably, $J_{0,\mathrm{Cl}}$ decreases from $m=6$ on (note that we plot $1-J_0$) in contrast to the GHZ encoding. This suggests that we cannot hope for a similar stabilization effect like for the GHZ encoding.

\textit{Comparing with active error correction.}--- The five-qubit 1d cluster state with periodic boundary conditions can be used as a codeword for conducting an error correction scheme in the presence of white noise \cite{GraphErrorCorr}. In this paragraph, we discuss and compare active error correction with the passive stabilization effect gained by the GHZ encoding.

We therefore use  Eq.~(\ref{eq:1c}) with $m=5$ to encode the quantum information. The Hilbert space of this logical block exhibits the dimension 32. We divide this space into 16 two-dimensional subspaces $\pi_i$. The first one, $\pi_0$, is spanned by the codewords itself; namely, $\ket{\mathrm{Cl}^{+}_5}$ and $\ket{\mathrm{Cl}^{-}_5}$. The remaining 15 subspaces are generated by applying to $\pi_0$ all possible one-particle errors (i.e., with $\sigma_{x=1}$, $\sigma_{y=2}$, and $\sigma_{z=3}$:  $\pi_{3(k-1)+i} = \sigma_i^{(k)}\pi_0 \sigma_i^{(k)}$, $k = 1,\dots,5$, and $i=1,2,3$). These 15 spaces represent all possible one-particle errors that may occur.

The procedure for the error correction is now the following. We start with a pure state that lies within $\pi_0$. Next, this state is subject to a noise process. For a moment, we only allow one error to occur. After some interaction time, we perform a measurement that allows us to distinguish between the 16 subspaces. If the measurement result indicates that we have projected onto $\pi_0$, we know that we have projected out all parts of the density matrix with ranges in the other subspaces $\pi_{i>0}$. We have therefore ``reinitialized'' the state. On the other hand, if we project onto one of the ``erroneous'' subspaces, say $\pi_4$, we can reverse the error by applying the same operation on the density matrix that maps $\pi_0$ to $\pi_4$, which is $\sigma_x^{(2)}$. Hence we have ``corrected'' the error that arose. Taking into account errors on more than one particle, we certainly cannot correct those. Therefore, the error correction works well if we encounter interactions with the environments that are so weak that the appearance of one-particle errors is much more probable than multiparticle ones.

\begin{figure}[htbp]
\centerline{\includegraphics[width=\columnwidth]{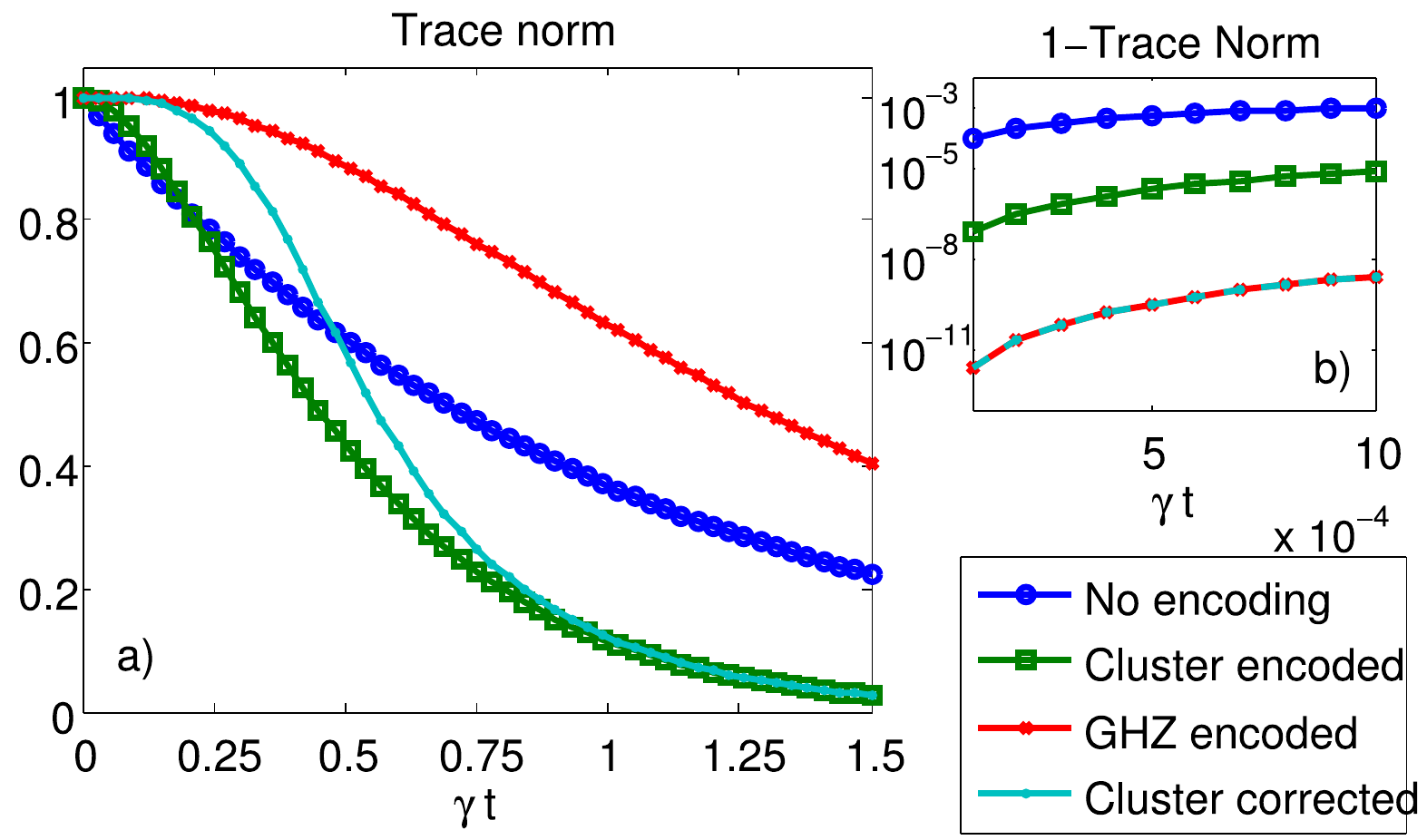}}
\caption[]{\label{fig:clusterCorr} (Color online) (a) We compare the trace norm of the off-diagonal elements of one block of a logical GHZ state for different situations: No encoding, GHZ encoding and cluster encoding with and without error correction [Eq.~(\ref{eq:28})]. Interestingly, the corrected GHZ behaves like the C-GHZ state for short times and like the uncorrected cluster encoding for larger times. (b) The same data for very short times. The coincidence of GHZ encoding and corrected cluster encoding is obvious.}
\end{figure}

In order to compare this scheme with the GHZ encoding, we focus on a more specific situation than just the norm of the off-diagonal elements. The reason is that simulating a measurement needs the full density operator instead of just some elements. To be in line with the present paper, we consider a logical GHZ state of  Eq.~(\ref{eq:2a}) with the encoding of Eq.~(\ref{eq:1c}), where we actively correct with the above-described protocol. We then compare the interference terms of the C-GHZ with $m=5$ with those of the logical GHZ with cluster encoding plus error correction. The measurement outcomes influence strongly the fidelity of the corrected state. For logical blocks in the $\pi_0$ subspace, the resulting norm of the off-diagonal element is more stable than for the C-GHZ state. Other measurement outcomes (corresponding to the occurrence of certain errors), however, lead to more unstable norms. We consider the convex combinations of all possible states after the measurement, weighted with the probabilities $w_i$ for the respective measurement outcomes $i$. The quantity we plot in Fig.~\ref{fig:clusterCorr} is the trace norm of
\begin{equation}
\label{eq:28}1/n
\sum_{i=0}^{15}w_i\pi_i \mathcal{D}_t
\left(\ketbra{\mathrm{Cl}_5^{+}}{\mathrm{Cl}_5^{-}}\right)\pi_i
\end{equation}
with $w_i = \mathrm{Tr}\left[\pi_i \mathcal{D}_t \left(\ketbrad{\mathrm{Cl}_5^{+}}\right)\right]$ and
\begin{equation}
n = \mathrm{Tr}\left[ \sum_{i=0}^{15}w_i\pi_i \mathcal{D}_t
\left(\ketbrad{\mathrm{Cl}_5^{+}}\right)\right].\label{eq:29}
\end{equation}
Comparing the trace norm of the interference terms of the corrected GHZ state and of the C-GHZ state, we find that, for small interaction times, they coincide, while for larger times the norm of the corrected case approaches the norm of the uncorrected cluster encoding, which drops below the standard GHZ state (i.e., $m=1$; see Fig.~\ref{fig:clusterCorr}). Notice that for larger times $t$ (i.e., small $p$), error correction no longer works properly as the probability for (uncorrectable) errors on two or more qubits increases. This is the regime where the stability of the encoded, corrected GHZ state falls below the stability of the uncoded GHZ state. It is important to stress that the corrected GHZ state does differ from the C-GHZ state in other properties (e.g., genuine multipartite entanglement). This comes from the fact that the error correction projects the state back to a smaller subspace, which is necessary to fulfill the conditions for genuine multipartite entanglement, see Sec.~\ref{sec:genu-mult-entangl}. On the other side, the C-GHZ obeys contributions in the whole Hilbert space which results in a relatively large trace norm but is not sufficient to guarantee GME.

\subsection{Cluster-GHZ state}
\label{sec:cluster-ghz}
The standard GHZ state exhibits interesting properties and applications (e.g. parameter estimation \cite{metrology,Huelga}), but is unstable under noise. In contrast, the 1D cluster state shows, to some extent, stability under decoherence \cite{ClusterStable,ConcatEC}. It is an interesting question whether the combination of both states leads to a new quantum state that is useful for parameter estimation while being more stable in the presence of environmental interactions. This was the motivation in Ref.~\cite{ClusterGHZ}, where the usefulness for parameter estimation of the quantum state
\begin{equation}
\label{eq:30}
\ket{\mathrm{CG}} = \frac{1}{\sqrt{2}}\left( \ket{\mathrm{Cl}_N^{+}} +  \ket{\mathrm{Cl}_N^{-}} \right)
\end{equation}
was investigated. This state is a superposition of two cluster states [see  Eq.~(\ref{eq:1c})]. We call $\ket{\mathrm{CG}}$ cluster-GHZ state in the following.

The cluster-GHZ state can be seen as an extension of the idea to encode the logical blocks by  Eq.~(\ref{eq:1c}). This leads us to the following investigations on other quantum properties besides the usability in the context of quantum metrology, which was done in Ref.~\cite{ClusterGHZ}. Here, we focus on the norm of the off-diagonal elements under white noise $\mathcal{D}_t\left( \ketbra{\mathrm{Cl}_N^{+}}{\mathrm{Cl}_N^{-}} \right)$.

\begin{figure}[htbp]
\centerline{\includegraphics[width=1.1\columnwidth]{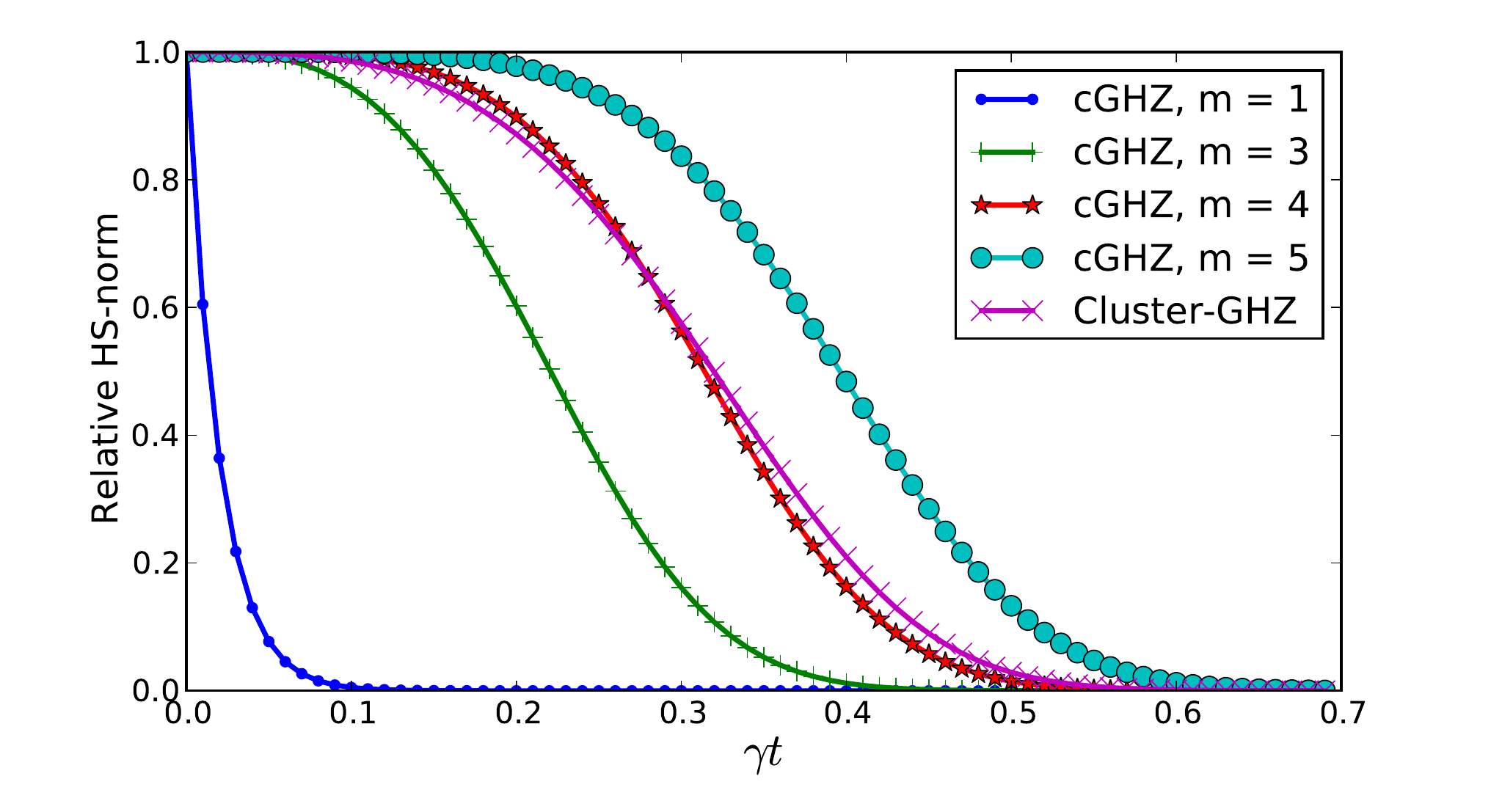}}
\caption[]{\label{fig:cl-ghz} (Color online) Relative HS norm of cluster-GHZ state [Eq.~(\ref{eq:31})] and several C-GHZ states [$\tilde{I}_0^N$ from Eq.~(\ref{eq:35})] are calculated for $N=100$ using the matrix product state (MPS) formalism, see Appendix A. We observe that the cluster-GHZ state behaves similarly as the C-GHZ state with $m=4$.}
\end{figure}

For large system sizes $N$, we cannot calculate the trace norm efficiently. In contrast, due to the simple structure of  Eq.~(\ref{eq:30}) in terms of tensor networks (see  Ref.~\cite{MPS} and references therein), we can easily calculate the HS norm (see Sec.~\ref{sec:hilbert-schmidt-norm}). We superficially review the concept of tensor networks in appendix \ref{sec:short-review-tensor} to give an idea how to compute the results presented here. The relative HS norm of the interference terms under white noise for the cluster-GHZ state
\begin{equation}
\label{eq:31}
\lVert \mathcal{D}_t\left( \ketbra{\mathrm{Cl}_N^{+}}{\mathrm{Cl}_N^{-}} \right) \rVert_2/\lVert \mathcal{D}_t\left( \ketbrad{\mathrm{Cl}_N^{+}} \right) \rVert_2
\end{equation}
is compared with those of several C-GHZ states [Eq.~(\ref{eq:35})]; see Fig.~\ref{fig:cl-ghz} for a fixed system size $N$ (note that $m=1$ represents the standard GHZ state). We emphasize that the total number of particles equals $N$ for the cluster-GHZ state, while is multiplied by $m$ in the case of the C-GHZ state. We observe that the cluster-GHZ state shows a similar behavior as the C-GHZ state for $m=4$. We see numerically that this is independent of the system size $N$.

In addition to this study we also calculate the distillability of the cluster-GHZ under white noise, as we have done for the C-GHZ in Sec.~\ref{sec:dist-prop-c}. This is interesting, since the 1D cluster state itself is distillable with a similar protocol, independent of the system size $N$ \cite{ClusterStable,ConcatEC}. For the cluster-GHZ state we get a different result. We consider a specific distillation protocol, where we measure all but two particles and are left with a two-body state $\rho_2$.
We then analyze the distillability of this resulting two-qubit state. If we measure every particle in the $\sigma_z$ basis, the quantum correlations of $\rho_2$ that give rise to distillable entanglement exclusively come from the off-diagonal element $\mathcal{D}_t\left( \ketbra{\mathrm{Cl}_N^{+}}{\mathrm{Cl}_N^{-}} \right)$. It turns out that these correlations rapidly vanish with $p^{N-4}$, which is of the same order as for the GHZ state. This means that with this specific protocol, the cluster-GHZ state is as poorly distillable as the GHZ state is.

It is, however, clear that we can use the distillability of the 1D cluster state itself in order to generate maximal entanglement out of the cluster-GHZ state. We simply have to perform measurements in the stabilizer basis, which project the cluster-GHZ state in one of the two branches, leading to a noisy 1D cluster state. From that, we can use the results of Refs.~\cite{ClusterStable,ConcatEC} for noisy 1D cluster states to show multipartite distillability.

\subsection{Concatenation of the encoding}
\label{sec:conc-encod}

Here we study encoded GHZ states [Eq.~(\ref{eq:2a})], where the logical codewords are obtained by a concatenated application of the encoding. This is similar to error correction, using concatenated quantum codes, however, without active intervention. A first-level encoding consists of replacing each physical qubit by a logical qubit, which itself is composed of $m$ physical qubits. At the second level, each of these physical qubits is again replaced by a logical qubit consisting of $m$ physical qubits, leading to a total of $m^2$ qubits for each logical block. In Ref.~\cite{ConcatEC} it has been shown that cluster states as codewords lead to an exponentially fast stabilization using error correction schemes \cite{Grassl02,QECC4}. Here we would like to investigate the effect of the concatenation for the GHZ and the cluster encoding on the relative HS norm without any active intervention. We take a logical block of 25 qubits and calculate the relative HS norm for four different logical encodings: the GHZ (\ref{eq:1}) and the cluster (\ref{eq:1c}) encoding with $m=25$ each and a second-level concatenated encoding for an $m=5$ GHZ and cluster encoding. In the case of the concatenated cluster encoding with tensor network techniques, the computation is sketched in Appendix \ref{sec:short-review-tensor}. Figure \ref{fig:concat} shows for the cluster encoding that the concatenation increases the stability which is in accordance with Ref.~\cite{ConcatEC} and Fig.~\ref{fig:ClusterCod1}, where we see that for the trace norm of the interference terms the increasing of the logical cell does not lead to a higher stability for the cluster encoding. For the GHZ encoding we observe the opposite: Here, the concatenation leads to more instability, reflecting that the GHZ encoding is optimal.

\begin{figure}[htbp]
\centerline{\includegraphics[width=1.1\columnwidth]{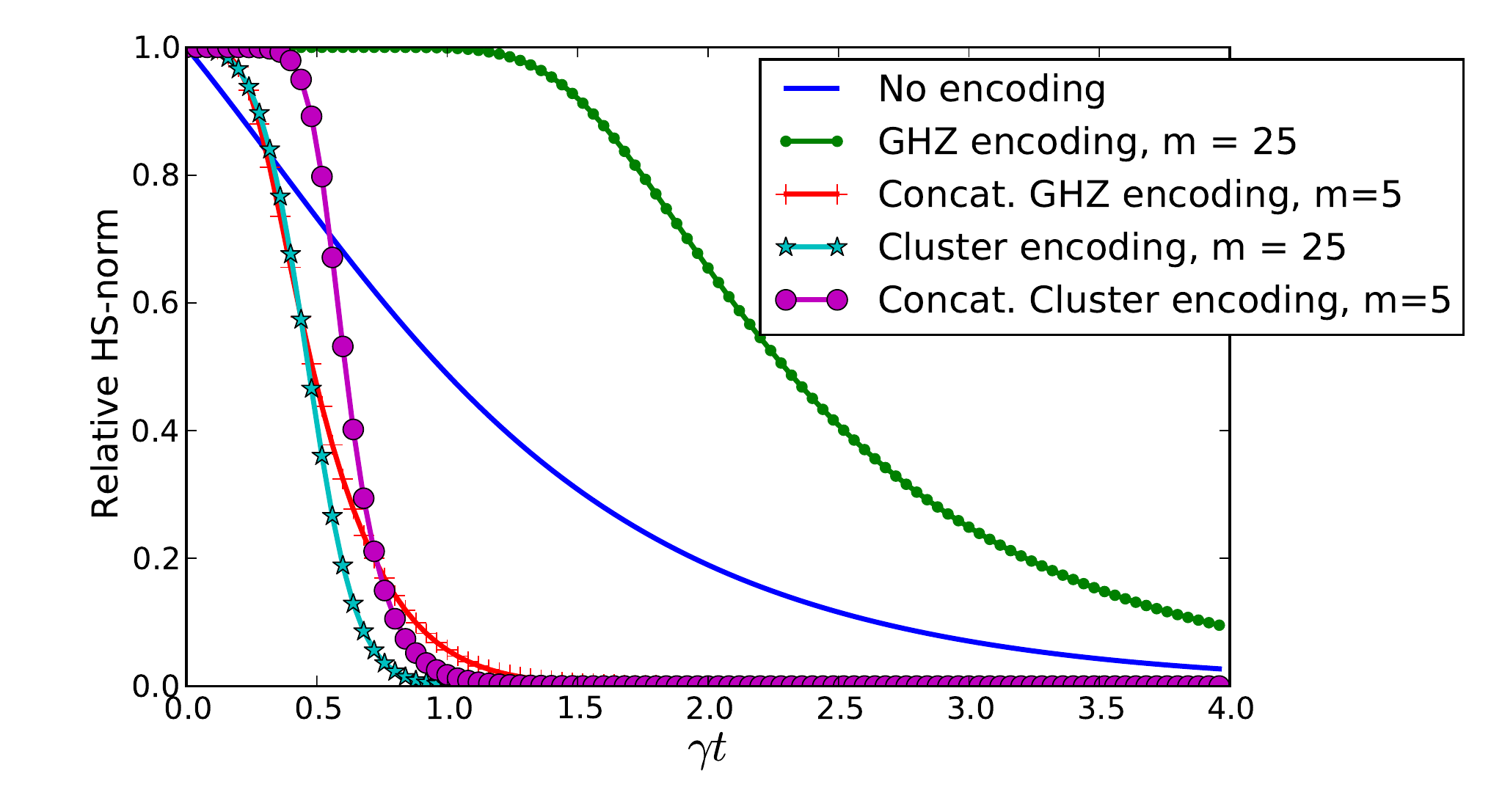}}
\caption[]{\label{fig:concat} (Color online) For fixed size of logical block (25 qubits), we compute relative HS norm for four different encodings (see text).}
\end{figure}

\subsection{Searching numerically for a stable encoding}
\label{sec:search-numer-stable}

\begin{figure}[htbp]
\centerline{\includegraphics[width=\columnwidth]{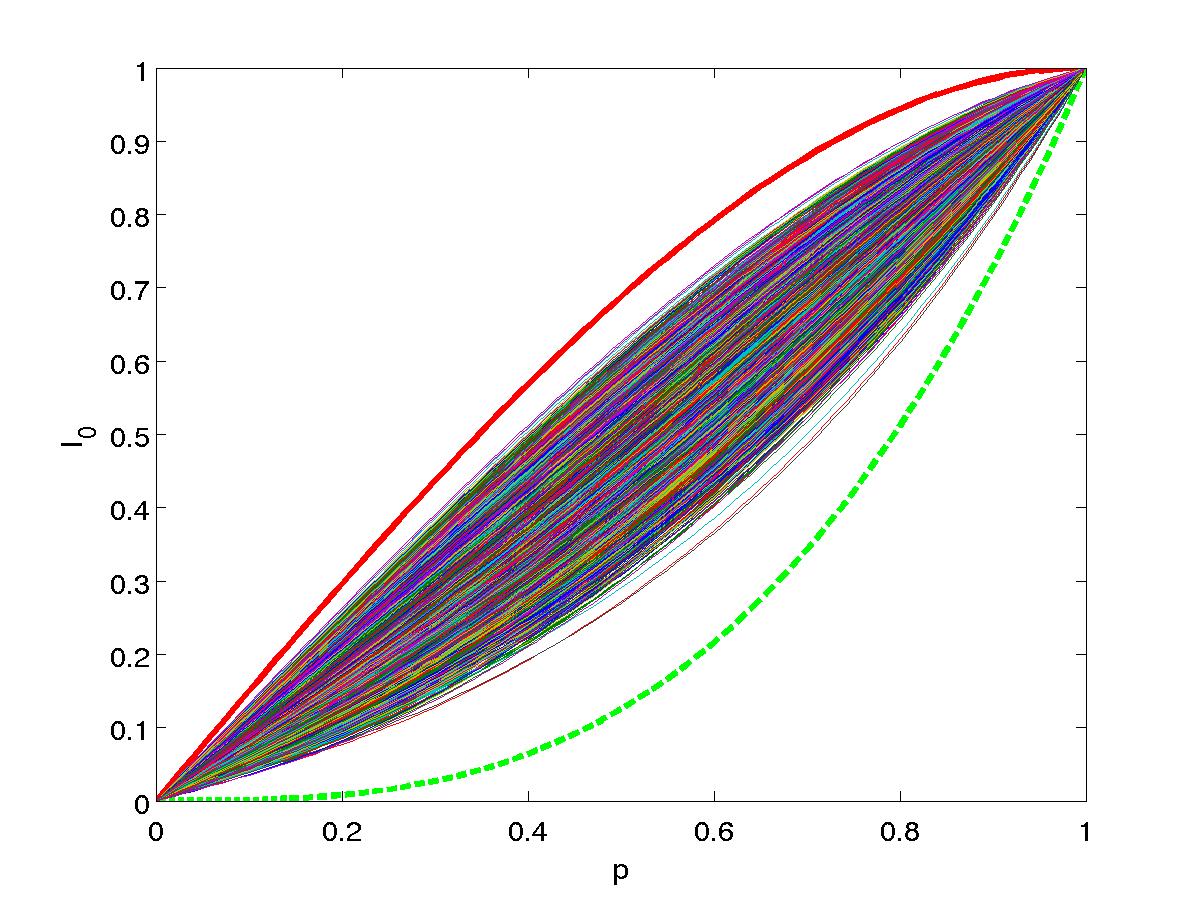}}
\caption[]{\label{fig:rand} (Color online) Numerical search for robust coding with three qubits. $10^5$ pairs of randomly generated, orthogonal states were subject to white noise. The trace norm of the off-diagonal element were calculated for different values of the noise parameter $p$. The thick red line corresponds to the GHZ encoding (\ref{eq:1}); the dashed green line to the ``worst scenario'' $\ket{\psi_0} = \ket{0}^{\otimes m}$ and $\ket{\psi_1} = \ket{1}^{\otimes m}$.}\end{figure}

We also investigate possible alternatives to the presented GHZ encoding by means of choosing generic state vectors of the Hilbert space $\mathcal{H}^{\otimes m}$ and (numerically) searching for stable encodings. For $m=3$, we generated randomly orthogonal pairs $\ket{\psi_0}$, $\ket{\psi_1}$ of states distributed uniformly by the Haar measure over the set of three-qubit states. We again consider GHZ states of the form $(|\psi_0\rangle^{\otimes N} + |\psi_1\rangle^{\otimes N})/\sqrt{2}$. It is sufficient to analyze only one logical block. We calculate the trace norm of the off-diagonal element under the influence of white noise $\lVert \mathcal{D}_t (\ketbra{\psi_0}{\psi_1})\rVert_1 $. Fig.~\ref{fig:rand} presents the results. None of the sampled $10^5$ encodings could exhibit a similar stability as the GHZ encoding, especially for short times ($p$ close to one). The GHZ encoding is the only one in this graph that obeys a vanishing first derivative for $t=0$. It seems that the product state encoding (leading to a standard GHZ state), and the GHZ state encoding (leading to a C-GHZ state) represent the extreme cases with respect to stability.

\subsection{Optimality and relation to active quantum error correction}
\label{sec:optim-relat-active}

In the previous sections we compared the passive GHZ encoding (\ref{eq:1}) and the resulting C-GHZ states with a number of different encodings, including an optimal 5-qubit code that can correct one arbitrary error, a concatenated code of this form, concatenated repetition codes, as well of encodings using random codewords. This includes both different passive encodings and active error-correction schemes. Perhaps surprisingly, we have identified the GHZ encoding to be optimal with respect to stability of coherences.

The observed stability does not seem to be directly related to properties of quantum error correcting codes, as the code we consider is actually a variant of a simple repetition code only capable of dealing with restricted kinds of errors. Quantum error correction codes are designed to protect quantum information in the presence of (local) noise and decoherence with help of repeated active error syndrome readout and correction. This implies also a certain robustness of entanglement properties of encoded systems; for example, distillability or negativity, even without active intervention. However, on the one hand, we also consider other interesting quantities such as the norm of coherences or index $q$, for which it is a priori not obvious whether they are equally well preserved by other encodings. In fact, we show that, regarding robustness of coherences, a passive GHZ encoding is superior to active quantum error correction with various codes.
On the other hand, error-correction codes are designed and optimized under the condition that active intervention is available and general quantum information should be preserved. As a result, many of these codes can only deal with a small amount of noise before an active intervention is required. This can be seen, for example, in the stringent error thresholds when using concatenated CSS codes for quantum communication \cite{KnillLaflamme96}, but also by the fact that concatenation decreases the effective noise at the next logical level only if the initial noise is small enough - about 10\% (see \cite{ConcatEC}). We find that a passive GHZ encoding allows us to preserve entanglement far beyond these limits (e.g., we find a nonzero negativity for up to 50\% of local white noise, see Fig.~\ref{fig:lifetime}). How well quantum features such as entanglement are preserved by general CSS codes (also in parameter regimes for which these codes are actually not designed) is an interesting open question.

\section{Experimental realization in ion trap setups}
\label{sec:exper-real-ion}

In addition to theoretical considerations, it is of fundamental interest to explore the experimental feasibility to generate a C-GHZ state in the laboratory. We assume a number of single-quantum systems (``particles'') where we can attach to every particle two orthogonal states in the corresponding Hilbert space. The reduction to this two-dimensional subspace is then called a qubit. Next, it needs access to entangling operations. Among several possibilities, the current tools developed in ion trap setups \cite{IonBlatts} appear to be suitable for an efficient creation of a C-GHZ state.

First, we will mention the available operations in a standard ion trap setup. On the one hand, we have the multipartite M\o lmer-S\o rensen (MS) gate \cite{MS,volckmar}
\begin{equation}
  \label{eqGen:2}
  \begin{split}
    U_n(\xi) &= \prod_{k=1}^{n-1}\prod_{l=k+1}^{n} U_{kl}(\xi)\\
    U_{kl}(\xi)&=\exp(i \xi \sigma_x^{(k)}\otimes \sigma_x^{(l)}).
  \end{split}
\end{equation}
In addition, we can use local rotations on the $z$ axis on parts of the chain
\begin{equation}
  \label{eqGen:3}
  Z(G) = \exp(i \pi/2 \sum_{k\in G}\sigma_z^{(k)}).
\end{equation}
The MS gate is capable of generating entanglement, since it is a product of two-body phase-gates. By induction, one can show that $U_n(\pi/2)$ transforms the state $\ket{0}^{\otimes n}$ into a GHZ state in a certain basis: the eigenbasis of $\sigma_z$, if $n$ is even; that of $\sigma_y$, if $n$ is odd:
\begin{subequations}\label{eqGen:5}
  \begin{align}
\label{eqGen:5a}
  U_n\left(\frac{\pi}{2}\right) \ket{0}^{\otimes n}&= \frac{e^{i\alpha}}{\sqrt{2}}\left(\ket{0}^{\otimes n}\pm                                                               i\ket{1}^{\otimes n}\right) \text{, n even,}\\
      \label{eqGen:5b}
  U_n\left(\frac{\pi}{2}\right) \ket{0}^{\otimes n}&= \frac{e^{i\beta}}{\sqrt{2}}\left(\ket{+}_{y}^{\otimes n}+
      \ket{-}_{y}^{\otimes n}\right) \text{, n odd.}
    \end{align}
\end{subequations}
The relative phase in  Eq.~(\ref{eqGen:5a}) equals $\pi/2$ or $-\pi/2$, depending on whether $n/2$ is odd or even, respectively.

The generation of the C-GHZ state by means of the MS gate can be drastically eased if we are able to use the MS gate only for a subgroup of qubits. Then we can simply apply this gate to any logical block in order to generate $\ket{0_L}^{\otimes N}\equiv\ket{\mathrm{GHZ}_m}^{\otimes N}$ out of $\ket{0}^{\otimes Nm}$. From here, we perform a global MS $\pi/2$-pulse to generate the full C-GHZ state.

In actual realizations, it is difficult to achieve MS gates that only involve certain subsets of particles, as this would involve precise focusing of the laser beams on the corresponding subset of ions. MS gates that act on all particles simultaneously, are much easier to realize. We hence consider only those gates in the following.  We therefore have to find a sequence of the gates (\ref{eqGen:2}) and (\ref{eqGen:3}) such that $\ket{0}^{\otimes Nm}$ is mapped on the state $\ket{\mathrm{GHZ}_m}^{\otimes N}$. Our main tool is the observation that
\begin{equation}
U_n(\xi)Z(\{j\}) = Z(\{j\}) \prod_{\substack{k<l\\ k,l \neq j}} U_{kl}(\xi) \prod_{\substack{k<l\\ k\, \mathrm{or}\, l =j}}U_{kl}(-\xi),\label{eqGen:4}
\end{equation}
that is, the phases for gates involving the site $j$ get a negative sign compared to the other pairs \cite{volckmar}. If we have two sites $G=\{i,j\}$, where we rotate before operating the MS gate, the phase of $U_{ij}(\xi)$ keeps its positive sign.

\begin{figure}[htbp]
\centerline{\includegraphics[width=.8\columnwidth]{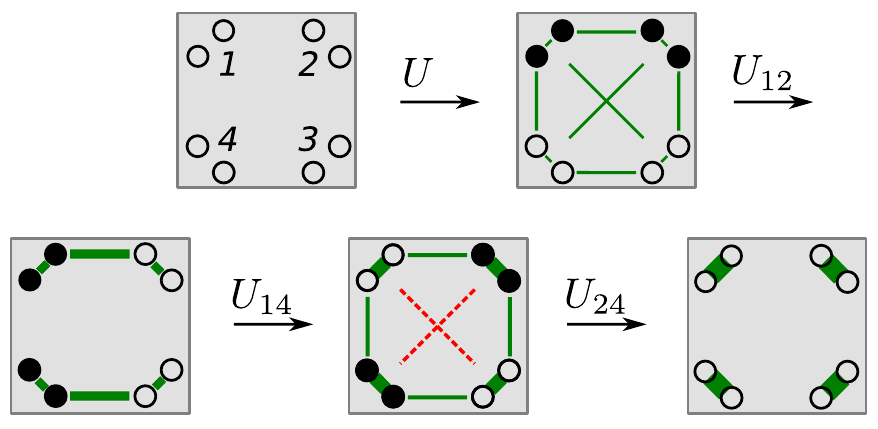}}
\caption[]{\label{fig:generating42} (Color online) On the generation of four Bell pairs out of a product state with the MS and the Z gate. The overall sequence reads $U Z_1 Z_4 U Z_1Z_{3} U Z_1Z_2U \ket{0}^{\otimes 8}$, where $Z_i \equiv Z(\{i_1,i_{2}\})$  and $U=U_8(\pi/8)$. The black-filled qubits mark the local rotations for the next step. The solid green lines correspond to a positive phase, the dashed red lines correspond to negative phases between a qubit pair. The thickness of the lines indicates the total number of applications of phase gates.}
\end{figure}

As an example we take $N=4$ and $m=2$ (see Fig.~\ref{fig:generating42}).
We apply the gate $U_8(\pi/8)$ four times. The total pulse length of the MS gate is therefore $\pi/2$. The three local rotations in between are chosen such that the phases add up to zero, except within the four pairs in the corners where we always have positive phases.

We show now that if $N$ is a power of two, we are able to generate an $N$-fold tensor product of GHZ states with $N$ applications of $U_{Nm}[\pi/(2N)]$ and $N-1$ local rotations of groups of $Nm/2$ ions. The explicit construction is an inductive one. Suppose we have a sequence of gates such that we are able to generate the state  $[U_{2m}(\pi/2)\ket{0}^{\otimes 2m}]^{\otimes N/2}$. This equals $\ket{\mathrm{GHZ}_{2m}}^{\otimes N/2}$ up to local unitaries. The fundamental blocks of this sequence are of the form $\tilde{U}_{i}(G) = U_{Nm}(\pi/N)\,Z(G_{i})$. $G_{i}$ is the empty set or includes half of the qubits. We now use the same sequence with half pulse duration [namely $\xi=\pi/(2N)$] acting on $\ket{0}^{\otimes Nm}$. After that, we repeat the sequence with modified $\tilde{U}_{i}(G)\rightarrow\tilde{U}_{i}(G)\,Z(H)$. $H$ is a fixed set and contains of every $2m$-group half of the ions.

The effective phases of the first and the second sequence between the groups of size $2m$ are zero. While for the first sequence we have the overall positive phase $\pi/4$ between the splitting $m:m$ within each group, the second sequence establishes exactly the opposite phase $-\pi/4$ and hence cancels the action of the first one. The phases within a $m$-group are constantly positive and add up to $\pi/2$. In fact, Fig.~\ref{fig:generating42} can be read in this manner. First we have two groups of size four (upper and lower half). The first two actions $U$ and $U\,Z_{1}\,Z_2$ lead to a state similar to $[U_4(\pi/4)\ket{0}^{\otimes 4}]^{\otimes 2}$. The last two steps are a repetition of the first two with $H = \{1,4\}$. [Notice that $Z(G)^2 = e^{i\alpha} \mathbbm{1}$.] Note that with the same arguments one can show that, if we have a sequence for $N$ groups, there is always a sequence of doubled length to achieve an $(N+1)$-fold tensor product of GHZ states.

To generate the full C-GHZ state, we put our findings together. For even $Nm$, we apply the MS gate in the first place, $U_{Nm}(\pi/2) \ket{0}^{\otimes Nm} \rightarrow 1/\sqrt{2}(\ket{0}^{\otimes Nm} + \ket{1}^{\otimes Nm})$ \cite{NoteGen}. Next we use the sequence discussed above, which accomplishes $\ket{0}^{\otimes Nm} \rightarrow \ket{0_L}^{\otimes N}$. The same sequence also implements $\ket{1}^{\otimes Nm} \rightarrow \ket{1_L}^{\otimes N}$, respectively up to the same unitary transformations. This finally leads to the generation of a C-GHZ of $N$ blocks and block size $m$. Notice that the complexity of the procedure depends only linearly on $N$ and is independent of the block size $m$.

If we associate $\xi$ with a dimensionless operation time, we see that the temporal cost for the MS gate is proportional to the minimal value. We need in total an interaction time of $\pi$, while the minimal value equals $\pi/2$. What changes with $N$ is the number of applications of the MS gate, which grows linearly. The number of local rotations as well as the total interaction time grows linearly with $N$. For systems up to $N=8$, we can show by a numerical search the optimality of our protocol. The assumptions for this search are a fixed pulse length of the local rotations and intermediate MS pulses of arbitrary length. We test all possible pulse sequences on their ability to perform $\ket{0}^{\otimes Nm}\rightarrow \ket{0_L}^{\otimes N}$ and count the number of pulses. For $N=2$ we need at least two, for $N=3,4$ four and for $N=5,\dots,8$ we need eight MS pulses with intermediate local rotations.

As a final remark, we want to stress that, even though the C-GHZ state is clearly superior to the standard GHZ state for large system sizes, already a few particles it suffices to show in proof-of-principle experiments the extended lifetime of entanglement. We can use stochastic local operations and classical communication (SLOCC) operations in order to reveal, for example, the negativity in the C-GHZ state. In this case, we simply have to project every logical qubit into the subspace spanned by $\ket{0_L}$ and $\ket{1_L}$. For $N = 4, m = 2$ and reasonable $p$ this can be done with a probability between 0.25 and 0.5. The projection concentrates the entanglement and leaves a four-qubit state that can be analyzed easily in the experiment. Notice that no actual projection onto the corresponding subspace is required, but it is sufficient to consider only postselected single-qubit measurements.

\section{Conclusion and Outlook}

\label{sec:conclusion}
In this paper, we have investigated encoded macroscopic quantum superpositions. We have analyzed in detail several quantum features of such states when subjected to local decoherence processes, including the decay of correlations (interference terms), the distillability properties, the negativity, the index $q$ and genuine multipartite entanglement. We have found generally valid bounds for interference terms, negativity and index $q$, all of which show that an exponential decay with the system size $N$ is unavoidable. However, these bounds also show that the decay can in principle be (exponentially) stabilized by increasing the size $m$ of the logical blocks. We have identified a particular interesting encoding, which was introduced in Ref.\cite{cGHZ}, where logical blocks are formed by orthogonal GHZ states, leading to the so-called concatenated GHZ states. For these states, we have indeed shown explicitly an exponential stabilization of coherence terms, negativity, distillability as well as the index $q$ with increasing block-size $m$. For $m = O(\log N)$ the decay can be effectively frozen.

A comparison with different encodings, including the usage of codewords of error correcting codes (without active error correction), concatenated encodings and choice of random states indicates that the GHZ encoding is in fact optimal for the stability of the trace norm of the interference terms. We have also pointed out that it is, in principle, possible to efficiently generate such C-GHZ states in an ion trap setup using present day technology. Similar approaches might be found for other setups. Interestingly, already moderate system sizes of $N=4$ and $m=2$ would suffice to observe stabilization effects.

We believe that the improved stability of the C-GHZ states as compared to standard GHZ states provides an important tool toward the generation of large-scale quantum superpositions, where distinct quantum effects might be observed at a mesoscopic or even macroscopic scale. Applications of such states in the context of quantum metrology seems very promising (see Ref. \cite{cGHZ} where the use of noisy C-GHZ states for parameter estimation is discussed). Parameter estimation using encoded quantum states, also in scenarios where noise and interactions simultaneously affect the system will be discussed in a future presentation.

\section*{Acknowledgments} The research was funded by the Austrian Science Fund (FWF):  P20748-N16, SFB F40-FoQus F4012-N16 and the European Union (NAMEQUAM).

\appendix

\section{Short review on tensor network techniques}
\label{sec:short-review-tensor}

We give a short overview on the concept of tensor network representations and computations, which is used in Sec.~\ref{sec:alternatives}. The tools to calculate the relative HS norm of the interference terms are sketched. For more extended explanations we refer to the literature (e.g., \cite{MPS,GroundStates,DMRG,MPOconstr}).

A quantum state of an $N$ qubit system consists of $2^N$ parameters. A generic state of this Hilbert space cannot be described efficiently. This is due to the complex structure of the correlations among the particles. In contrast, a product state can be described efficiently, since the number of independent parameters scales linearly with $N$. The idea of tensor network states is to find an efficient way of describing a quantum state that obeys some correlations among the particles which are either well structured --like in the case of the GHZ state-- or of short range (which imposes a geometry onto the problem). The same is true for general linear operators on this space. As an instance consider a linear operator $O: \mathbbm{C}^{2 \otimes N}\rightarrow \mathbbm{C}^{2 \otimes N}$ decomposed in a local basis $\sigma$
\begin{equation}
\label{eq:43}
O = \sum_{\substack{i_1, \dots, i_N=0\\j_1, \dots, j_N=0}}^{1}c^{j_{1},\dots,j_N}_{i_1,\dots,i_N} \sigma^{j_{1}}_{i_{1}} \otimes \dots \otimes \sigma^{j_{N}}_{i_{N}}.
\end{equation}

\begin{figure}[htbp]
\centerline{\includegraphics[width=0.85\columnwidth]{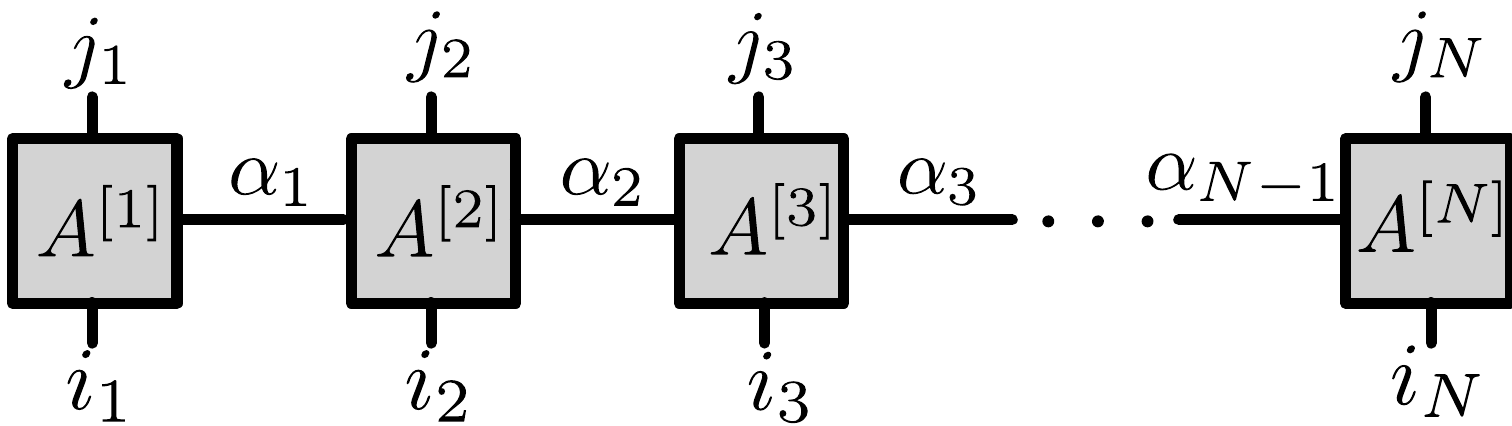}}
\caption[]{\label{fig:MPO} Matrix product operator representation. An operator $O$ acting on $N$ particles is decomposed into $N$ low-rank tensors $A^{[k]}$. Each tensor has two physical indices (input $i_k$, output $j_k$) and one or two virtual indices $\alpha_{k-1}, \alpha_k$ which are summed over.}
\end{figure}

The high-rank tensor $c^{j_{1},\dots,j_N}_{i_1,\dots,i_N}$ can always be decomposed into a large number of low-rank tensors. Typically, we assign one tensor to every particle. The geometry of the problem gives rise to the network structure. Most popular are one-dimensional geometries, since there one cannot only represent weakly correlated states and operators efficiently, but one can as well compute expectations values in an efficient manner. In one-dimensional networks, representations of states and operators are usually called matrix product states (MPSs) and matrix product operators (MPOs), respectively \cite{MPS,Schadschneider,MPS2}. For $O$, the MPO representation is achieved by decomposing the tensor $c$:
\begin{equation}
\label{eq:44}
  c^{j_{1},\dots,j_N}_{i_1,\dots,i_N}=\sum_{\alpha_1, \dots, \alpha_{N-1}=1}^{D}A^{[1]j_{1}}_{i_1 \alpha_1}A^{[2]j_{2}}_{\alpha_1 i_2 \alpha_2}\dots A^{[N]j_{N}}_{\alpha_{N-1} i_N}.
\end{equation}
It is common to depict decompositions like in Eq.~(\ref{eq:44}) graphically. Tensors are drawn as boxes with legs, where each leg represent an index. The summation over a common index
of two tensors is indicated by the connection of the respective
legs to tensors; see Fig.~\ref{fig:MPO} as a representation of Eq.~(\ref{eq:44}).

The indices $\alpha_k$ are called virtual indices. Their dimensions, also known as bond dimensions, are crucial for the efficiency of the representation. With MPS and MPO representations one can compute similarly as with the standard vector and matrix representations. Efficient MPS and MPO representations can be added and multiplied efficiently. In general, the bond dimension of the sum is the sum of the bond dimensions of the addends, the bond dimension of the product is the product of the bond dimensions of the two factors. In particular, every linear operation on a MPO can be efficiently computed. The full power of this representation method can be found in numerical studies on ground-state searches of one-dimensional Hamiltonians \cite{GroundStates}, which are related to the successful density matrix renormalization group (DMRG) algorithms \cite{DMRG}.

\textit{Tools to compute the results in Sec.~\ref{sec:alternatives}}--- In order to calculate the HS norm of the interference terms, we generate the MPS representation of the orthogonal states, build the corresponding matrix element, apply the noise process and calculate the HS norm of this operator. In the following, this is described in some detail. The procedure for the HS norm of the diagonal elements is completely analogous.

First we generate the 1D cluster state which is used in Sec.~\ref{sec:cluster-ghz}. The pattern of applying phase gates $C$ on a product state imposes a one-dimensional geometry like in Fig.~\ref{fig:MPO}, but here we form a ring of tensors. We start with the product state $\ket{+}^{\otimes N}$ or $\ket{-}^{\otimes N}$ (trivial bond dimension one) for $\ket{\mathrm{Cl}^{+}_N}$ or $\ket{\mathrm{Cl}^{+}_N}$ respectively
and apply to every two neighbors a phase gate $C$. This operator has a MPO representation with bond dimension two for the link between the two particle it acts on and trivial bond dimension everywhere else (see also Ref.~\cite{MPOconstr} for general explanation how to find those descriptions). After all $C$ have been applied, the cluster state representation exhibits a constant bond dimension two between any two neighbors.

\begin{figure}[tb]
\centerline{\includegraphics[width=\columnwidth]{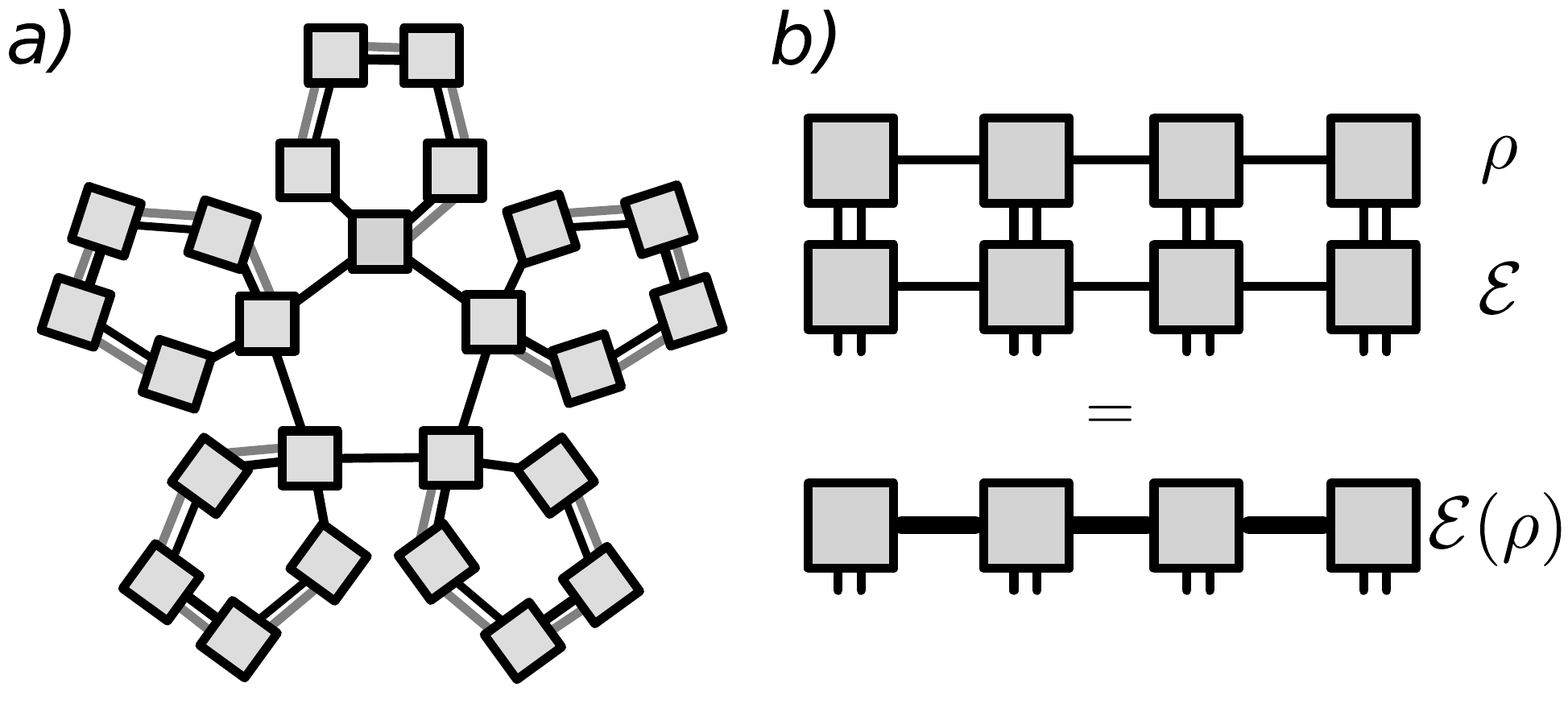}}
\caption[]{\label{fig:mpo2} (a) On the generation of concatenated cluster encoding with 25 particles (see text). Here, the black lines represent applications of phase gates, while gray lines refer to controlled-\textsc{not} gates. The physical indices, which every tensor obeys, are not drawn. (b) On the action of a general superoperator on a density operator. Every tensor of the superoperator representations exhibits four physical indices; two of them (the ``input'' indices) are summed over with the physical indices of the tensors of the density operator representation. The thicker line of the $\mathcal{E}(\rho)$ representation indicates that the bond dimension is the product of the bond dimensions of the $\rho$ and $\mathcal{E}$ representations.}
\end{figure}

The concatenated structure of cluster states in Sec.~\ref{sec:conc-encod} leads to a different geometry depicted in Fig.~\ref{fig:mpo2}~(a). We start with a cluster ring of five qubits as before. Next we attach to every qubit four more qubits. We first apply a controlled-\textsc{not} operation, where the qubit within the inner ring controls the four other qubits simultaneously (bond dimension two). Then local Hadamard operations act on all qubits and finally phase gates between all neighbors on the outer ring are placed. In total, the phase gates and the controlled-\textsc{not} operation together account for a bond dimension four on the outer rings; the inner ring exhibits two-dimensional bonds.

The matrix element $\ketbra{\mathrm{Cl}^{+}_N}{\mathrm{Cl}^{-}_N}$ is gained by directly multiplying the MPS representations of $\ket{\mathrm{Cl}^{+}_N}$ with $\bra{\mathrm{Cl}^{-}_N}$. This is simply obtained by an outer product of the $i^{\mathrm{th}}$ tensor in the representation of $\ket{\mathrm{Cl}^{+}_N}$ with the (complex conjugated) $i^{\mathrm{th}}$ tensor in the representation of $\ket{\mathrm{Cl}^{-}_N}$. The new tensors have to be reshaped in order to exhibit the standard form of two physical and two virtual indices. This leads to a squaring of the bond dimension.

Now we calculate the action of the cp map onto the operator, $\mathcal{E}\left(\ketbra{\mathrm{Cl}^{+}_N}{\mathrm{Cl}^{-}_N}  \right)$. To this aim, one needs to generalize the above tensor network description to linear operations that map operators on operators (sometimes also called superoperators). Such a tensor network description of cp maps is straightforward. As before, the tensor network representation of a general superoperator consists of $N$ tensors, each now of rank six (five at the boundary); four physical indices (two ``input'' and two ``output'' indices) and two virtual indices to both neighbors. If we want to apply the superoperator to a density matrix, the physical indices of the tensors of density operator have to be ``connected'' to the input indices of the tensors of the superoperator; connecting indices is equivalent to summing over those indices. Fig.~\ref{fig:mpo2} (b) shows a sketch of this procedure. The presence of uncorrelated cp maps lead to an application of single-particle superoperators. That means that the matrix product representation of $\mathcal{E}$ has bond dimension one and the complexity of the noisy state in terms of the bond dimension is not altered. Hence we are able to follow the time evolution exactly. This provides us with a powerful tool to study the influence of noise and decoherence on multipartite entangled states consisting of a large number of particles.

The last ingredient is to calculate the Hilbert-Schmidt norm $\lVert A \rVert_2 = \sqrt{\mathrm{Tr}A A^{\dagger}}$ of a MPO. We therefore multiply the MPO with its hermite conjugate (squared bond dimension) and trace over the product. The trace of $O$ in Eq.~(\ref{eq:43}) is performed by summing all indices $i_k$ with their respective counterparts $j_k$. The maximum bond dimension is 16 in the case of standard cluster states and 256 for the concatenated structure.



\begin{thebibliography}{99}
\bibitem[]{NoteConnectionMacroMeasure}In fact, in Ref.~\cite{DSC} the decay rate of off-diagonal elements was used to assess an ``effective size'' to certain pure macroscopic superposition states. In view of the results of \cite{cGHZ} and the findings of this article (Sec.~\ref{sec:c-ghz-state}), it however seems that in the case of encoded superpositions, such an approach is generically not applicable.
\bibitem[]{NoteLocalMeas} Here, the phrase local refers to the situation where the experimenter has access to the whole block and can perform any desired operation on a single block.
\bibitem[]{NoteOrder} The definitions we use here are $f(N) = O(N^x) :\Leftrightarrow \lim_{N\rightarrow \infty}f(N)/N^x > 0$ and exists. We also need $f(N) = o(N^x) :\Leftrightarrow \lim_{N\rightarrow \infty}f(N)/N^x = 0$.
\bibitem[]{NoteAssumption} To show that $\lVert C\left[\mathcal{D}(\ketbrad{\psi_0})\right]\rVert_{1} = o(N^2)$, it is sufficient to show that the index $q$ does not increase under uncorrelated cp maps. For the depolarization channel this can be seen easily since the Kraus operators of $\mathcal{D}$ are unitary operators and the action of $\mathcal{D}$ on a density operator $\rho$ leads to a convex combination of differently rotated density operators. But since the cp map acts uncorrelated, the rotations are local and therefore cannot increase the index $q$. This statement is true for all so-called unital cp maps \cite{NiCh00}. For general cp maps it is still open whether the index $q$ can increase under their action.
\bibitem[]{NoteGen}For odd $Nm$, we begin with $\ket{+}^{\otimes Nm}$ and the action of the MS gate in the $y$-basis.
\end{thebibliography}
\end{document}